\documentclass[%
aps, 
 prf,
 longbibliography,
preprint,
 amsmath,amssymb,
]{revtex4-1}

\usepackage{graphicx}
\usepackage{amsfonts}
\usepackage{amssymb}
\usepackage{float}
\usepackage{latexsym}
\usepackage{amsmath}
\usepackage{placeins}
\usepackage{subcaption}
\usepackage[euler]{textgreek}
\usepackage{textcomp}
\usepackage{siunitx}

\usepackage[table]{xcolor}

\newcommand{\eqr}[1]{(\ref{eq:#1})}
\newcommand{\od}[2]{\frac{d #1}{d #2}}
\newcommand{\pd}[2]{\frac{\partial #1}{\partial #2}}

\newcommand{\eg}{\textit{e.g.}}
\newcommand{\bs}[1]{\boldsymbol{#1}}
\newcommand{\etal}{\textit{et al.}}
\newcommand{\ie}{\textit{i.e.}}

\graphicspath{{./figures//}}

\begin{document}

\title{Natural oscillations of a sessile drop on flat surfaces with mobile contact lines}

\author{Jordan Sakakeeny}
\author{Yue Ling}%
 \email{Stanley\_Ling@baylor.edu}
\affiliation{Department of Mechanical Engineering, Baylor University, Waco, TX, USA}




\date{\today}

\begin{abstract}
Oscillation of sessile drops is important to many applications. In the present study, the natural oscillation of a sessile drop on flat surfaces with free contact lines (FCL) is investigated through numerical and theoretical analysis. The FCL condition represents a limit of contact line mobility, i.e. the contact angle remains constant when the contact line moves. In the numerical simulation, the interfaces are captured by the volume-of-fluid method and the contact angle at the boundary is specified using the height-function method. The oscillation frequencies for sessile drops with FCL are mainly controlled by the contact angle and the Bond number and a parametric study is carried out to characterize their effects on the frequencies for the first and high-order modes. Particular attention is paid to the frequency of the first mode, since it is usually the dominant mode. An inviscid theoretical model for the first mode is developed. The model yields an explicit expression for the first-mode frequency as a function of the contact angle and the Bond number, with all parameters involved fully determined by the equilibrium drop theory and the simulation. The predicted frequencies for a wide range of contact angles agree very well with the simulation results for small Bond numbers. The frequencies for both the first and high-order modes decrease with the contact angle and increase with the Bond number.  For the high-order modes, the frequencies for different modes generally scale with the Rayleigh frequencies. The scaling relation performs better for small Bond numbers and large contact angles. A simple model is proposed to predict the frequencies of high-order modes for large contact angles and a good agreement with the simulation results is observed. 
\end{abstract}

\maketitle


\section{Introduction}
Oscillation of sessile drops on flat supported surfaces is important to numerous applications, including drop shedding on condensation surfaces \cite{Yao_2017a}, growing crystal \cite{Strani_1984a}, water harvesting \cite{Dai_2018a}, among many others. In particular, recent studies have shown that oscillations of sessile drops, induced by acoustics or surface vibrations, can enhance the mobility of the drop \cite{Yao_2017a} or even cause the sessile drops to detach from the wall \cite{Boreyko_2009a}. When the forcing frequencies match with the natural frequencies of the drops, the excited oscillation amplitude will be maximized for a given forcing amplitude, due to the resonance effect \cite{Boreyko_2009a, Chang_2013b, Noblin_2004a}. Therefore, it is highly desirable to accurately predict the natural/resonance frequencies for sessile drops on different material surfaces.

The investigation of the natural oscillations of a free drop can be traced back to the pioneering work of Rayleigh \cite{Rayleigh_1879a}. Based on the assumptions of inviscid free-surface flows, the frequencies of infinitesimal-amplitude oscillations of a free drop were determined analytically. The study of Rayleigh was then extended by Lamb to include the azimuthal mode shapes and to incorporate the effect of ambient fluid \cite{Lamb_1932a}. The effect of the liquid viscosity can be characterized by the Ohnesorge number. For drops with large Ohnesorge number, the oscillation frequency is reduced due to the viscous effect \cite{Reid_1960a,Miller_1968a, Prosperetti_1980a, Basaran_1989a}. When the oscillation amplitude is large, the oscillation frequencies decrease {due to the nonlinear effect} \cite{Trinh_1982a,Tsamopoulos_1983a} and the effect of inter-mode coupling becomes significant \cite{Basaran_1992a}. The present study is primarily interested in small-amplitude oscillations of low-viscosity liquid drops. In such cases, the oscillation frequency $\omega$ perfectly scales with the capillary frequency $\omega_c$ and the normalized frequency $\omega/\omega_c$ is only a function of the mode number. 

For a sessile drop sitting on a flat surface, additional parameters arise due to the interaction between the drop and the surface \cite{Chang_2013b}. When the capillary and gravity forces are in balance, the drop is at its equilibrium state. The equilibrium shape of a sessile drop varies with the contact angle $\theta$ and the gravitational Bond number $Bo = \rho_l g R_d^2 / \sigma$, where $\rho_l, g, R_d$ and $\sigma$ are liquid density, gravity, volume-based drop radius, and surface tension, respectively. For example, sessile drops with small $\theta$ or large $Bo$ exhibit lens-like shapes, while those with large $\theta$ and small $Bo$ take drop-like shapes. As a result, the oscillation frequencies also vary with $\theta$ and $Bo$. The normalized frequencies $\omega/\omega_c$ for a sessile drop generally decrease as the contact angle $\theta$ increases \cite{Bisch_1982a, Strani_1984a, Basaran_1994a, Bostwick_2014a}, approaching the asymptotic limit at $\theta=\ang{180}$. In this limit, the constraint from the surface vanishes and the drop becomes spherical like a free drop. If gravity is absent, the oscillation frequencies for $\theta = \ang{180}$ reduce to the Rayleigh or Lamb frequencies for all modes \cite{Strani_1984a}. The oscillation frequencies are also influenced by the gravity effect, which is characterized by $Bo$. The numerical studies of Basaran and DePaoli \cite{Basaran_1994a} have shown that the frequencies of pendant drops (sessile drops with negative gravity) decrease with $Bo$.

In addition to $\theta$ and $Bo$, the contact-line mobility can also affect the oscillation dynamics for a sessile drop \cite{Noblin_2004a,Chang_2013b}. When contact-line hysteresis is present, the contact angle for a moving contact line is different from the equilibrium contact angle. The difference between the receding and advancing contact angles is often used to characterize the hysteresis effect. Though modeling moving contact lines in continuum mechanics remains an unresolved challenge \cite{Bonn_2009a}, some of the difficulties can be alleviated by focusing on the two asymptotic limits for the contact-line mobility: 1) the pinned contact line (PCL) limit, for which the contact line is fixed/pinned while the contact angle can vary to a large extent, and 2) the free contact line (FCL) limit, for which the contact line can move freely while the contact angle is fixed at its equilibrium value. The oscillation frequencies for general sessile drops with moving contact lines and non-zero contact-angle hysteresis will be bounded between these two limits. Bostwick and Steen \cite{Bostwick_2014a} showed that the natural frequency for PCL is generally higher than that for FCL for a given mode.

The oscillations of sessile drops with PCL were first studied experimentally by Bisch \etal \cite{Bisch_1982a}. The drop was placed on the top of a cylindrical pillar and, as a result, the contact line was pinned at the edge of the cylinder. The oscillation frequencies for different axisymmetric modes were measured. Strani and Sabetta \cite{Strani_1984a} solved the free-surface potential flow induced by the axisymmetric oscillations. In order to expand the potentials to spherical harmonic modes, the original constraint of a flat surface was replaced with a spherical bowl of radius equal to that of the equilibrium shape of the drop. The problem eventually reduced to an eigenvalue problem and was solved using the Green function method. The original inviscid theory has also been extended to incorporate the viscous effect \cite{Strani_1988a}. The first asymmetric oscillation mode (note that the first mode does not involve shape oscillation for a free drop) was identified and the predicted first-mode frequencies for different $\theta$ generally agreed well with the experiments of Bisch \etal \cite{Bisch_1982a}, though some discrepancy was observed for small $\theta$. 
Instead of constraining the drop using a spherical bowl, Bostwick and Steen \cite{Bostwick_2009a} only pinned the drop surface at the circle of contact and allowed the surface under the contact line to deform. This treatment is consistent with a double sessile-pendant drop system in a capillary switch \cite{Ramalingam_2012a}. When the constraint circle overlaps with a nodal line of the corresponding free drop, then there will be no influence to the oscillation modes and the natural frequencies \cite{Prosperetti_2012a}. 
{Bostwick and Steen have also identified the oscillation frequencies for sessile drops on flat surfaces through a linear inviscid stability analysis \cite{Bostwick_2014a}. The effect of gravity is ignored, so the equilibrium shapes of the sessile drops are spherical caps. Three contact-line conditions: PCL, FCL, and the Hocking condition (where the contact angle varies smoothly with the contact-line speed \cite{Davis_1980a,Hocking_1987a}) were considered. The predicted oscillation frequencies and mode shapes for the PCL condition have been validated by experiment \cite{Chang_2015a}. The inviscid theory of Bostwick and Steen for the FCL condition was validated only for the contact angle $\ang{90}$, for which 
the oscillations for the sessile drops become similar to those for a free drop with twice volume \cite{Bostwick_2014a}. Further examinations of their model for FCL at contact angles other than $\ang{90}$ are still required. }

While the above theoretical studies attempted to solve the potential flows exactly (with different boundary conditions at the contact line), simplified theoretical models have also been developed to predict the natural oscillation frequencies. Noblin \etal \cite{Noblin_2004a} proposed a model based on 1D capillary waves to predict the oscillation frequency for PCL as well as mobile contact lines. The model results agree reasonably well with the experimental results for low contact angles. Nevertheless, recent experiments by Yao \etal \cite{Yao_2017a} showed that the model is less accurate for sessile drops on superhydrophobic surfaces with PCL and large $\theta$.  Celestini and Kofman \cite{Celestini_2006a} have developed a theoretical model using a different approach to predict the frequency for the lateral oscillation (parallel to the surface) of a sessile drop with PCL. The key assumption in their model is that the shapes of the sessile drop at different oscillation phases are similar to the equilibrium shapes of the sessile drop when a body force of different magnitude and sign (positive or negative) is applied. The equilibrium shape is numerically solved to provide the quadratic relations between drop surface area and the centroid deviation for different contact angles. The model predictions agreed reasonably well with their own experimental data for $\theta=\ang{140}$. 

Though extensive efforts have been made regarding the oscillations of sessile drops, as comprehensively reviewed by Bostwick and Steen \cite{Bostwick_2015a}, it is clear that the effects of the contact angle, Bond number, and contact-line mobility on the oscillation frequency of sessile drops are still not fully understood. This is particularly true for sessile drops with FCL and finite $Bo$. In recent years, research advances have been made on superhydrophobic surfaces \cite{Rothstein_2010a} and slippery liquid-infused porous surfaces \cite{Wong_2011a} to reduce contact-angle hysteresis and to enhance drop mobility. The dynamics of contact line on these surfaces is close to the FCL limit. Understanding the oscillation of sessile drops with FCL is therefore important to predict the behavior of drops on such surfaces. The purpose of the present study is to systematically investigate the effects of contact angle and Bond number on the frequencies of axisymmetric oscillation modes for a sessile drop on flat surfaces with FCL. The azimuthal modes and the azimuthal contact-line instability \cite{Vukasinovic_2007a} are not considered in the present study. A combined computational and theoretical approach will be adopted. Particular attention is paid to the first oscillation mode, which is a unique feature for sessile (or pendant) drops and is generally the dominant mode. Axisymmetric, fully-resolved simulations are performed using the volume-of-fluid method for a wide range of $\theta$ and $Bo$. Furthermore, a new inviscid theoretical model for the first mode is developed in the present study. {The previous theoretical models of Strani and Sabetta \cite{Strani_1984a} and Bostwick and Steen \cite{Bostwick_2014a}  have assumed zero gravity and spherical-cap equilibrium shape. These assumptions are removed in the present model and the effect of the Bond number on the oscillation frequency is incorporated. An important feature of the new model is that the first-mode oscillation frequency can be expressed as an explicit function of $\theta$ and $Bo$, making it much more convenient to use, compared to previous theoretical models \cite{Strani_1984a, Bostwick_2014a}.}

The numerical methods and the simulation setup will be presented in Section \ref{sec:sim}. The theoretical model will be introduced in Section \ref{sec:theory}. The results for both the simulations and the theoretical model will be presented and discussed in Section \ref{sec:results}. Finally we will conclude the key findings in Section \ref{sec:conclusions}. 


\section{Simulation methods}
\label{sec:sim}
\subsection{Governing equations}
The one-fluid approach is employed to resolve the gas-liquid two-phase flows induced by the sessile drop oscillation. The liquid and the gas are treated as one fluid with material properties that change abruptly across the gas-liquid interface. The Navier-Stokes equations for incompressible flow with surface tension is given as
\begin{equation}
  \rho (\delta_t \mathbf{u} + \mathbf{u} \cdot \nabla \mathbf{u}) = -\nabla p + \nabla \cdot (2 \mu \mathbf{D}) + \sigma \kappa \delta_s \mathbf{n},
  \label{eq:NS1}
\end{equation}
\begin{equation}
  \nabla \cdot \mathbf{u} = 0,
  \label{eq:NS2}
\end{equation}
 where $\rho$, $\mu$, $\mathbf{u}$, and $p$ represent density, viscosity, velocity, and pressure, respectively. The strain-rate tensor is denoted by $\mathbf{D}$. The third term of the right-hand side of Eq.\ \eqr{NS1} is a singular term, with a Dirac distribution function $\delta_s$ localized on the interface, and it represents the surface tension. The surface tension coefficient is $\sigma$, and $\kappa$ and $\mathbf{n}$ are the local curvature and unit normal of the interface, respectively.

The volume fraction of liquid $C$ is introduced to distinguish the different phases; $C = 0$ and 1 for cells containing only air and water, respectively. The time evolution of $C$ satisfies the advection equation
\begin{equation}
  \delta_t C + \mathbf{u} \cdot \nabla C = 0.
  \label{eq:C1}
\end{equation}

The fluid density and viscosity are determined by 
\begin{align}
  \rho & = C \rho_l + (1 - C) \rho_g\,,
  \label{eq:density1}\\
  \mu &= C \mu_l + (1 - C) \mu_g\, ,
  \label{eq:viscosity1}
\end{align}
where the subscripts $g$ and $l$ correspond to the gas phase (air) and the liquid phase (water), respectively.

\subsection{Numerical methods}
The Navier-Stokes equations (Eqs.\ \eqr{NS1} and \eqr{NS2}) are solved by the open-source multiphase flow solver, \textit{Basilisk} \cite{basilisk}. The \textit{Basilisk} solver uses a finite-volume approach based on a projection method. A staggered-in-time discretization of the volume-fraction/density and pressure leads to a formally second-order-accurate time discretization. The advection equation for liquid volume fraction (Eq.\ \eqr{C1}) is solved using a geometric Volume-of-Fluid (VOF) method \cite{Scardovelli_1999a, Popinet_2009a}.  A {quadtree} spatial discretization is used, which gives a very important flexibility by allowing adaptive mesh refinement in user-defined regions. The height-function (HF) method is used to calculate the local interface curvature \cite{Popinet_2009a} and to specify the contact angle on the surface \cite{Afkhami_2009a}. Finally, a balanced-force surface tension discretization is used \cite{Francois_2006a,Popinet_2009a}. Validation studies for the numerical methods and the \textit{Basilisk} solver in solving a wide variety of interfacial multiphase flows can be found in previous studies, \eg, \cite{Zhang_2019b, Marcotte_2019a, Zhang_2020a, Mostert_2020a}.

\subsection{Simulation setup}
\subsubsection{Computational domain and boundary conditions}
\begin{figure}
 \centering{\includegraphics[trim=0 0.5in 0 0in,clip, width=0.5\textwidth]{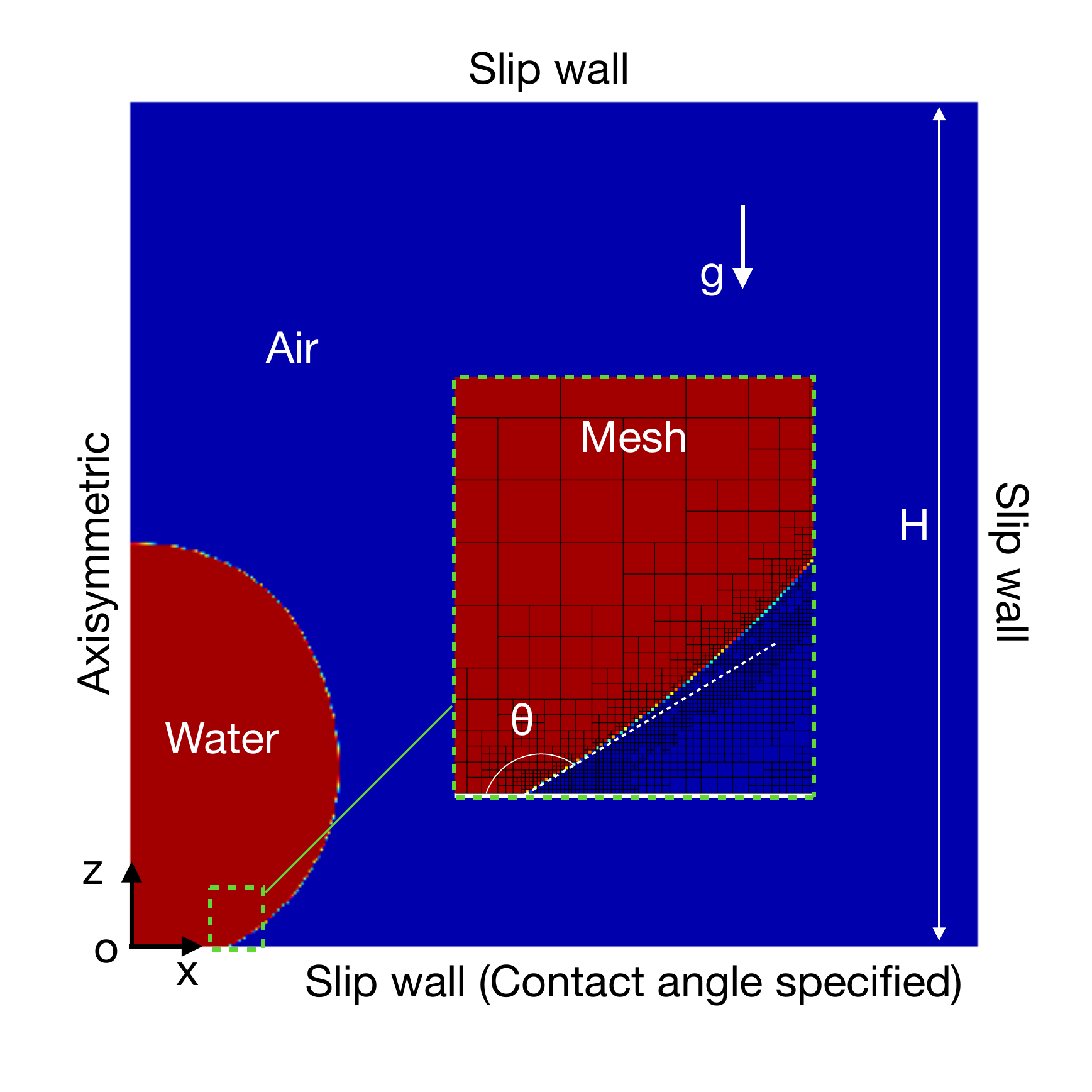}}
 \caption{Simulation setup and mesh. The contact angle $\theta$ is specified at the bottom wall.}
 \label{fig:SimSetup}
\end{figure}

{
For a spherical harmonic oscillation mode $Y_n^m$, the azimuthal modes ($m\neq0$) share the same frequency of the zonal mode ($m=0$) for the same degree $n$. Since the focus of the present study is to predict the oscillation frequency, only the axisymmetric zonal modes will be simulated. }
The computational domain for the 2D-axisymmetric simulations is shown in Fig. \ref{fig:SimSetup}. The length of the square domain edge, $H$, is about four times the drop radius. 

The axisymmetric boundary condition is applied on the left surface. The top and right surfaces are slip walls. The sessile drop is sitting on the bottom wall, where the contact angle is specified as a user-defined constant, using the height function method \cite{Afkhami_2009a}. For a given contact angle, the orientation of the interface in the cell containing the contact line is fixed, as a result, the change of the liquid volume fraction $C$ in that cell will induce a slip of the contact line. In the FCL limit, there is no resistance to the contact line motion, so the contact angle remains constant. Therefore, to be consistent, the bottom wall is set as a slip wall. Tests for no-slip conditions at the  bottom wall have also been performed and it has been verified that the oscillation frequencies for slip and no-slip bottom walls are almost the same. 

\subsubsection{Physical Parameters}
\label{sec:parameters}
\begin{figure}
 \centering{\includegraphics[trim=0 0.in 0 0in,clip, width=1\textwidth]{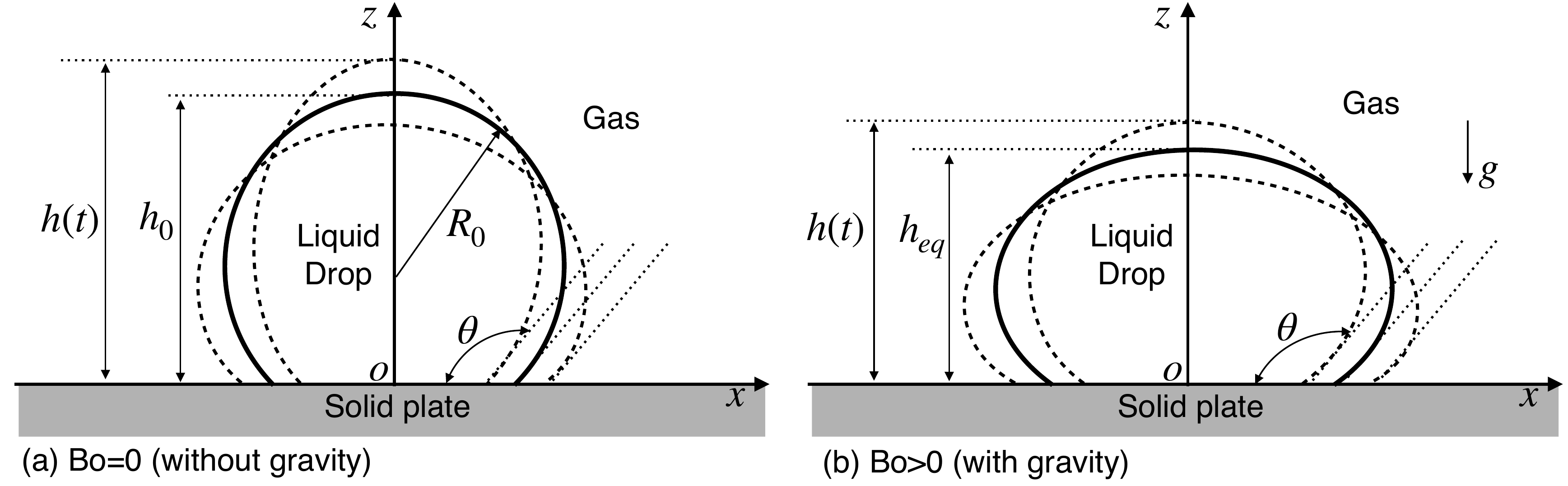}}
 \caption{Schematic of a sessile drop oscillating on a slippery surface with constant contact angle, (a) without and (b) with the action of gravity. The solid lines represent the equilibrium state. }
 \label{fig:schematic}
\end{figure}

The liquid and gas here are taken to be water and air, respectively. The fluid properties are listed in Table \ref{tab:physParam}. The contact angle on the surface, $\theta$, is varied from $\ang{50}$ to 150\textdegree. 
{The wide range of contact angles $\theta$ considered here is sufficient to cover different common materials used in dropwise condensation \cite{Yao_2017a, Dai_2018a} and former experimental and theoretical studies of sessile drops \cite{Vukasinovic_2007a, Bostwick_2014a, Chang_2015a}. For $\theta>\ang{150}$, the contact area between the drop and the surface is small, and drop oscillations tend to cause the drop to jump off from the surface  \cite{Boreyko_2009a}. Though drop jumping is of great interest, it is out of the scope of the present study.}
The volume of the drop is kept constant, $V_d=65.45 \mu L$, across all cases, for which the volume-based radius $R_d=(3V_d/4\pi)^{1/3}=2.5$mm. Since the viscosity of water is low, the Ohnesorge number of the millimeter-size drop is quite small. The variation of Ohnesorge number due to the change of drop volume has little effect on the normalized oscillation frequency, $\omega/\omega_c$, where $\omega$ and $\omega_c$ are the drop oscillation and capillary frequencies, respectively. A parametric study for the drop volume for $Bo=0$ is presented in Appendix \ref{sec:AppendixVarVol}, affirming that the effect of $V_d$ on $\omega/\omega_c$ is small. The capillary frequency $\omega_c$ is defined as 
\begin{equation}
	\omega_c = \sqrt{\sigma / (\rho_l R_0^3)}\,,
\end{equation}
where $R_0$ is the radius of the spherical-cap sessile drop at the equilibrium state when gravity is absent, see Fig.\ \ref{fig:schematic}(a), and is a function of $V_d$ and $\theta$ as
\begin{align}
    R_0 & = \left(\frac{3V_d}{\pi (2 + cos \theta) (1-cos \theta)^2}\right)^{1/3} \,
    \label{eq:R0}
\end{align}
In other words, when gravity is absent, the oscillation frequency for low $Oh$ simply scales as $\omega \sim \omega_c$ or $\omega \sim V_d^{-1/2}$. For finite $g$, the effect of the drop volume on the normalized oscillation frequency is mainly reflected in $Bo$. In the present study a wide range of $Bo$ is considered by varying $g$ from zero to full gravity. As the radius for the sessile drop at equilibrium state is not a constant any more, see Fig.\ \ref{fig:schematic}(b), the volume-based radius, $R_d$, is used as the reference length scale. 

%

The values of the key dimensionless parameters, defined based on scaling variables $R_d$, $\rho_l$, and $\sigma$, are listed in Table \ref{tab:physParamNonDim}. The gas-to-liquid density and viscosity ratios are quite small, so the effect of the surrounding air on the drop oscillation is minimal. Furthermore, the small Ohnesorge number $Oh=0.00571$ indicates that the effect of liquid viscosity on the oscillation frequency is negligible. For the present problem, the two important parameters are the contact angle $\theta$ and the Bond number $Bo=\rho_l g R_d^2 / \sigma$ and the effect of which will be systematically investigated.  
Alternatively, the Bond and Ohnesorge numbers can also be defined based on $R_0$ as $Bo_0=\rho_l g R_0^2 / \sigma$ and $Oh_0=\mu_l / (\rho_l \sigma R_0)$, which will vary with $\theta$, even when $V_d$ is fixed.

{
The contact line motion is generally dissipative and will affect the oscillation dynamics, such as the decay rate of the oscillation amplitude. However, the influence of the dissipation on the oscillation frequency is small for low $Oh$ droplets. That is why the inviscid theories in previous studies \cite{Strani_1984a, Celestini_2006a, Bostwick_2014a} well predict the oscillation frequencies even for the PCL conditions, where the dissipative effect is significant. The present study is focused on the FCL condition, where the contact angle remains constant as the contact line moves. In this limit, there is no dissipation induced by the contact-line motion \cite{Davis_1980a, Bostwick_2014a}. 
}

\begin{table}
    \centering
    \begin{tabular}{c c c c c c c}
        \hline
        \begin{tabular}{@{}c@{}}$\rho_l$ \\ $(kg/m^3)$ \\ \end{tabular} & 
        \begin{tabular}{@{}c@{}}$\rho_g$ \\ $(kg/m^3)$ \\ \end{tabular} & 
        \begin{tabular}{@{}c@{}}$\mu_l$  \\ $(Pa\ s)$ \\ \end{tabular} &
        \begin{tabular}{@{}c@{}}$\mu_g$  \\ $(Pa\ s)$ \\ \end{tabular} & 
        \begin{tabular}{@{}c@{}}$\sigma$ \\ $(N/m)$ \\ \end{tabular} &
        \begin{tabular}{@{}c@{}}$V_{d}$ \\ $(\mu L)$ \\ \end{tabular} & 
        \begin{tabular}{@{}c@{}}$\theta$ \\ (degree) \\ \end{tabular} \\
        \hline
         $1000$ & $1.2$ & $1 \times 10^{-3}$ & $1 \times 10^{-5}$ & $0.07$ & $65.45$  & 50-150\\
         \hline
    \end{tabular}
    \caption{Physical parameters.}
\label{tab:physParam}
\end{table}
%

\begin{table*}
    \centering
    \begin{tabular}{c c c c c c}
        \hline
        r & $m$ & $Oh$ & $\theta$ & $Bo$\\
        \hline
        $\rho_g / \rho_l$ & $\mu_g / \mu_l$ & $\mu_l / (\rho_l \sigma R_d)$ & (degree) & $\rho_l g R_d^2 / \sigma$  \\
        \hline
         $0.0012$ & $0.017$ & 0.0057 & $50-150$ & 0-0.88\\
         \hline
    \end{tabular}
    \caption{Key dimensionless parameters.}
\label{tab:physParamNonDim}
\end{table*}

\subsubsection{Mesh resolution}
A quadtree mesh is used to discretize the domain. The local cell size is adapted based on the estimated discretization errors of the liquid volume fraction and the fluid velocity. The assessment of discretization error for each scalar is achieved through a wavelet transform \cite{Hooft_2018a}. If the estimated error is larger than the specified threshold, the mesh will be locally refined, or vice versa. For the present simulation, the normalized error thresholds for both the volume fraction and velocity are set as 0.001. A representative snapshot of the  mesh close to to the contact line is shown in Fig.~\ref{fig:SimSetup}. It can be seen that the adaptation threshold is sufficient to specify a high enough mesh resolution to resolve the drop surfaces. 

The minimum cell size in the quadtree adaptive mesh is controlled by the maximum refinement level ($L$), \ie, $\Delta x_{\min}=H/2^L$. The mesh for $L=10$ is used in the present simulation, which corresponds to $R_0/\Delta x_{\min} \approx 256$. A grid refinement study varying $L=8$ to 11 has also been performed to fully confirm the results are mesh independent.  

\subsection{Initial Conditions}
Two different approaches have been used to perturb the equilibrium shape and to initiate the drop oscillation. The equilibrium shape of the sessile drop can be obtained by solving an ODE system, given in Appendix \ref{sec:equilibrium}. 
The first method (denoted as IC1) is to increase the gravity magnitude for a short duration. The equilibrium shape for a given $Bo$ (or $g$) is specified at $t=0$, then the $g$ is increased by $g_{pert}$ for $t\le t_{pert}$, where the perturbation gravity $g_{pert}=4.9$ m/s$^2$ and the perturbation time $t_{pert}\omega_c=$0.56 are used in the simulations. Due to the increased gravity, the drop will be pushed down and deviate from the equilibrium shape. Once the gravity returns to the original value (for $t>t_{pert}$), the drop shape will deform back toward the equilibrium shape. Though all oscillation modes will be initiated to some extent, the first mode ($n=1$) dominates in the spectrum. 

The second method (denoted as IC2)  is employed mainly to better capture the high-order ($n>1$) modes. The initial drop contour is specified as 
\begin{equation}
    x(z,t=0) = x_{eq}(z) + x'(z)
    \label{ODE}
\end{equation}
where $x_{eq}(z)$ is the equilibrium drop contour and the perturbation $x'(z)$ takes  the form of sinusoidal functions,
 \begin{align}
	x'(z) = \chi \sum_{n=1}^{10} c_n  \sin\left(\frac{n\pi z}{h_{eq}}\right)
\label{perturbation}
\end{align}
where $h_{eq}$ represents the height of the sessile drop at the equilibrium state, see Fig.\ \ref{fig:schematic}(b).
The overall perturbation amplitude is controlled by the parameter $\chi$ and a small value, $\chi = 0.005$, is used here to guarantee the induced oscillations are linear and also to prevent the drop from jumping off of the surface for large contact angles. The coefficients $c_n$ in the perturbation represent the weights for the sinusoidal functions $\sin(n\pi z)$. The values for $c_n$ used here are $c_1=0$, $c_2=0.25$, $c_{3,4}=0.5$, and $c_n=1$ for $n\ge4$, which are biased toward to the higher wavenumber to ensure that the induced high-order modes contain sufficient initial energy and can be clearly identified in the frequency spectrum. The oscillation frequencies are not affected the specific values of $c_n$. It is noted that the sinusoidal functions are not the exact eigenfunctions for the drop shape oscillation, yet the IC2 method here is effective to trigger high-order oscillation modes. 

\subsection{Summary of simulation cases}
To systematically investigate the effects of contact angle ($\theta$) and Bond number ($Bo$) on the oscillation of a sessile drop, 11 different values of $\theta$ (from $\ang{50}$ to $\ang{150}$ with an increment of $\ang{10}$) and 11 different values of $Bo$ (from 0 to 0.88 with an increment of $0.088$). Therefore, a total of 121 cases are considered in the parametric study. 

For each combination of $\theta$ and $Bo$, the ODEs for the equilibrium drop theory are solved to provide the equilibrium shape. Furthermore, two methods (IC1 and IC2) have been used to initialize drop oscillations. As a result, a total of 242 simulations have been performed, in addition to the simulations for the validation cases and other tests shown in the appendices. 

The simulations 
were performed on the Baylor University cluster \emph{Kodiak} using 4 CPU cores (Intel E5-2695 V4). Each simulation case takes about  60 to 80 hours of CPU time to reach the time $t\omega_c\approx 113$. The simulation time is generally sufficient to measure the frequencies for all the oscillation modes. 

\begin{figure}%
    \centering
    \includegraphics[width=1\textwidth]{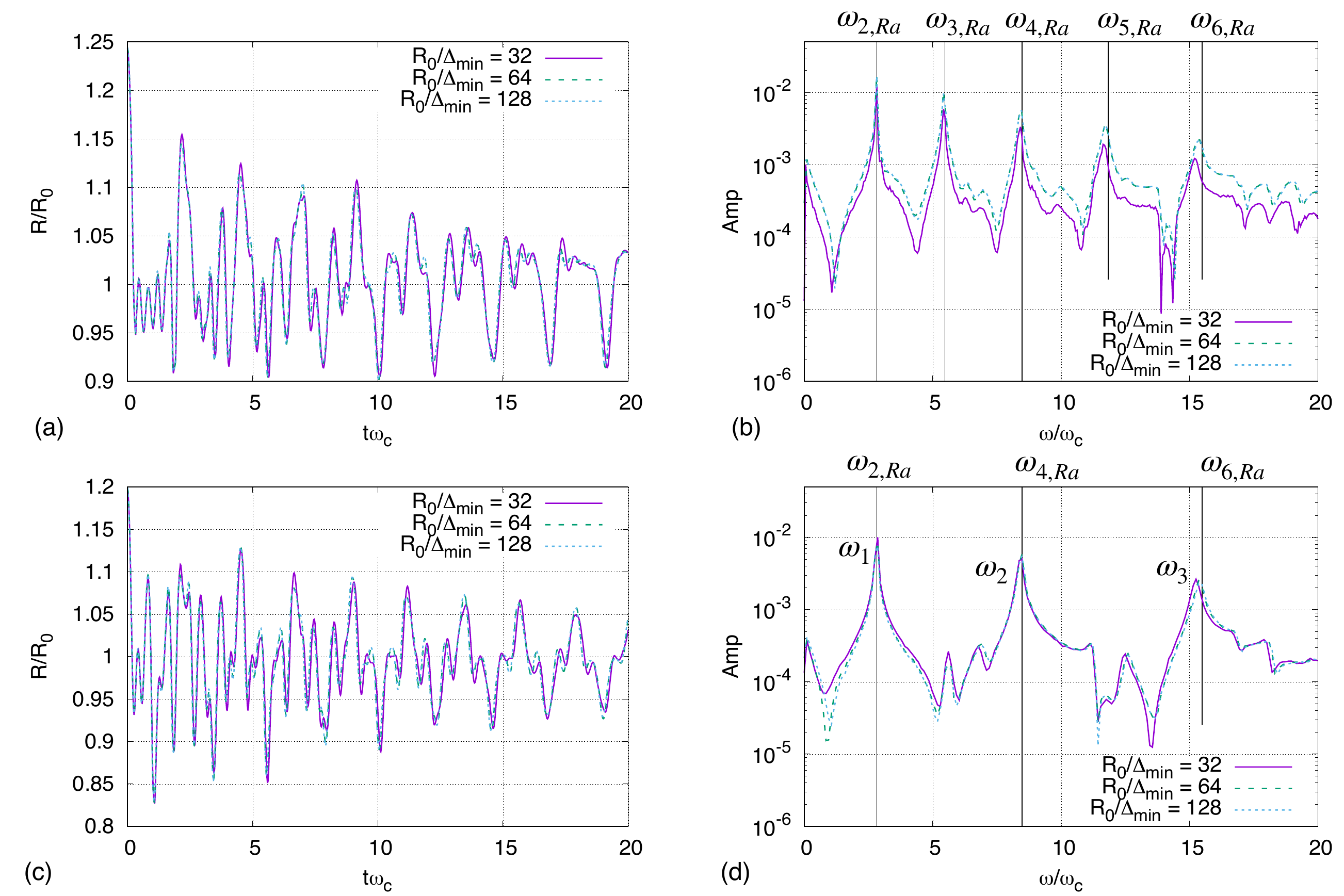}
    \caption{{Temporal evolution and frequency spectrum for the drop radius at $\phi=0$ for the natural oscillation for a free drop (a)-(b) and for a sessile drop with $\theta=\ang{90}$ (c)-(d). The simulation results for three different refinement levels are shown. The vertical lines in (b) and (d) indicates the Rayleigh frequencies.}}
    \label{fig:validation}%
\end{figure}

{
\subsection{Validation: oscillations of a free drop and a sessile drop with $\theta=\ang{90}$}
}
\label{sec:validation}
To validate the simulation approaches described above, the shape oscillation of a free drop without gravity is simulated. The same drop volume is used.  Three different maximum refinement levels have been tested, which correspond to $R_0/\Delta_{\min} = [32, 64,128]$, respectively. To initiate the shape oscillation, the radius of the drop at $t=0$ is perturbed using the Legendre polynomials, 
\begin{equation}
	\frac{R(\phi,t=0)}{R_d} =  \sum_{n=0}^{6}A_nP_n(\cos (\phi))\,,
	\label{eq:Ledendre}
\end{equation}
where $\phi$ is the colatitude and is taken to be zero at the top of the drop, $P_n$ is the Legendre polynomial of degree $n$, and $A_n$ is the corresponding Fourier-Legendre coefficient. Due to incompressibility,  the drop volume is fixed and $A_0=1$. The centroid location $z_c$ is fixed in time, and thus $A_1=0$. Here, the second to the sixth modes are considered and $A_n=0.05$ for $n=2$ to 6. The perturbation amplitudes are small, so the induced oscillation is expected to follow the linear theory of Rayleigh \cite{Rayleigh_1879a}. 
The drop radius at $\phi=0$, $R_{\phi=0,t}$, is measured from the top of the drop to the centroid. The temporal evolution and the frequency spectra of $R_{\phi=0,t}$ for three different meshes are shown in Fig. \ref{fig:validation}. The results for $R_0/\Delta_{\min} =64$ and 128 match very well, indicating the mesh resolution is sufficient to capture the shape oscillation. Furthermore, the frequencies for the second to the sixth modes can be clearly identified from the spectra, which match very well with the corresponding Rayleigh frequencies. 

{To validate that the simulation approaches in resolving the oscillations of sessile drops with FCL, a sessile drop with contact angle $\theta=\ang{90}$ is simulated, with three different mesh resolutions $R_0/\Delta_{\min} = 32, 64$, and 128. Again, the results for the different $\Delta_{\min}$ agree very well for both the temporal evolution of the drop radius and the frequency spectra, indicating the mesh used is also fine enough for sessile drop oscillations. The sessile drop has the same $R_0$ as the free drop. Due to the FCL and the slip-wall boundary condition on the solid surface, this sessile drop with $\theta=\ang{90}$ is similar to the upper half of the free drop. The oscillations of the sessile drop for the $n^{th}$ mode are similar to those of the free drop for the $(2n)^{th}$ mode. The oscillation frequencies for the sessile drop $\omega_{1,2,3}$ identified in the spectra are observed to agree very well with the Rayleigh frequencies $\omega_{2,4,6}$ for the free drop, see Fig.\ \ref{fig:validation}. This good agreement validates the present simulation approaches for sessile drop oscillations with FCL. The frequency of a  sessile drop with $\theta=\ang{90}$ has also been used for  model validation by Bostwick and Steen \cite{Bostwick_2014a}. 
}

\section{Theoretical model for the first mode}
\label{sec:theory}
The first oscillation mode is a unique feature for sessile drops \cite{Strani_1984a, Bostwick_2009a}. For a free drop, the first mode corresponds to a pure translation of the drop centroid and will not induce any shape oscillation. In contrast, for a sessile drop with mobile contact lines, when the centroid of the drop moves away from the equilibrium position, the radius of the drop must change correspondingly to satisfy mass conservation, accompanied by a shape oscillation. Furthermore, the first mode usually dominates in natural oscillation and thus is the easiest mode to be excited by external forcing to enhance drop mobility \cite{Boreyko_2009a}. This is due to the fact that the damping rate of an oscillation mode generally increases with mode number $n$ \cite{Lamb_1932a}, so the high-order modes always decay faster than the first mode. A theoretical model is developed in this section to better understand the first-mode oscillation dynamics and to predict the corresponding frequencies for different $\theta$ and $Bo$. 

In the model, the fluid motion is considered as inviscid and the effect of the surrounding air is ignored. {Consistent with the inviscid assumption, the solid surface is treated as a slip wall. As a result, the viscous singularity that occurs at the contact line on a no-slip boundary can be avoided \cite{Bostwick_2014a, Dussan_1979a}}. The oscillation amplitude is taken to be small. Consistent with simulations, the contact angle is taken to be constant when the contact line moves due to drop oscillation. Since there is no viscous dissipation, the total energy, $E_{tot}$, is conserved. The total energy is the sum of three contributions 
\begin{equation}
  E_{tot} = E_s + E_g + E_k
  \label{TotalEnergy}
\end{equation}
where $E_s$, $E_g$ and $E_k$ are the surface, potential, and kinetic energy, respectively. 

\subsubsection{Surface energy}
\begin{figure}
 \centering
 \includegraphics[width=0.6\textwidth]{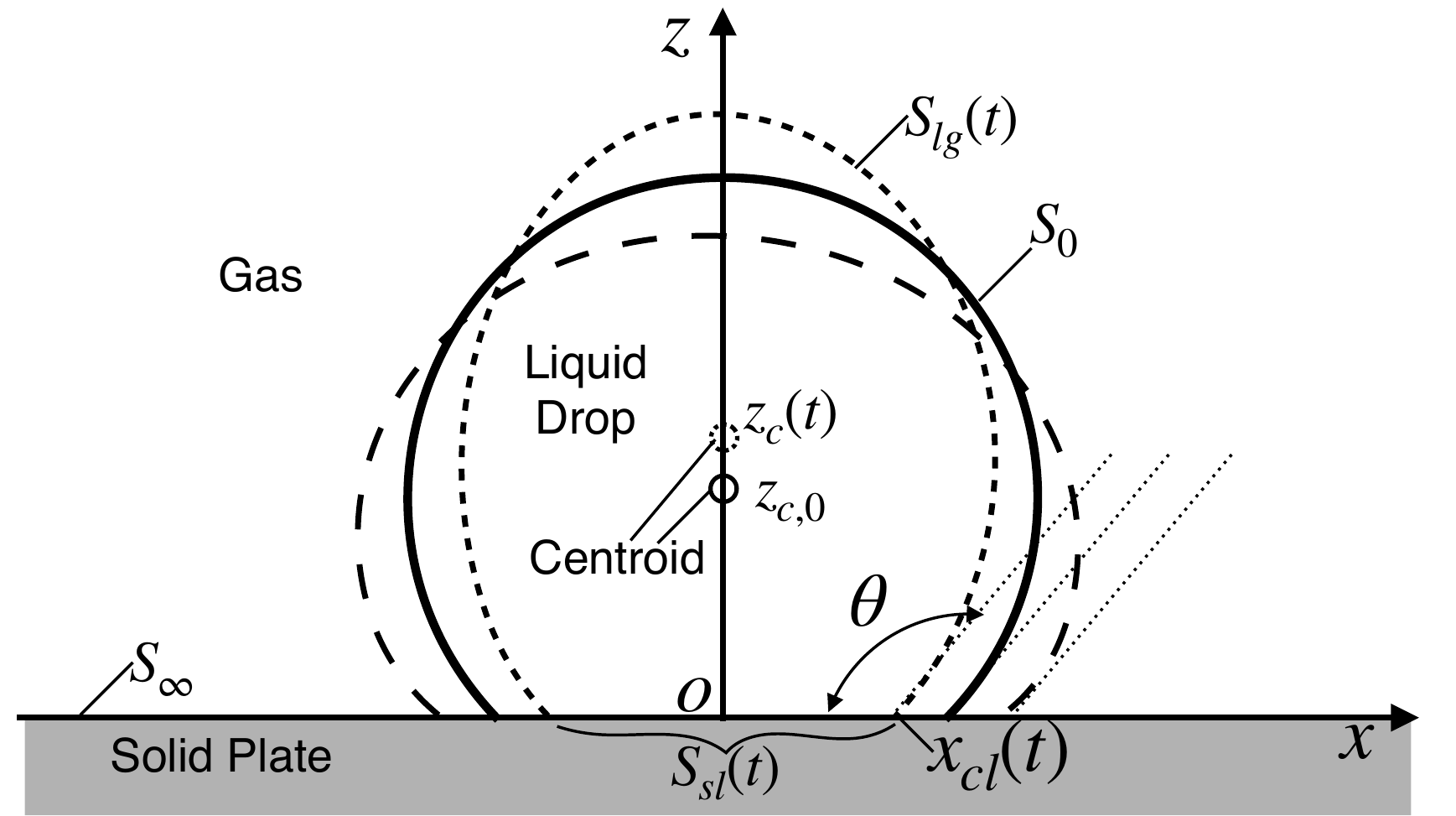}
 \caption{Schematics to show the temporal evolution of the drop centroid location, the areas for the liquid-gas surface, liquid-solid surface, and gas-solid surface, when a sessile drop oscillates on a slippery surface with a constant contact angle and no gravity.}
 \label{fig:Theoretical_DefDraw}
\end{figure}

The surface energy of a sessile drop at a given time is the sum of the energy contained in the liquid-gas, gas-solid, and liquid-solid surfaces, given as 
\begin{equation}
  E_{s}'(t) = \sigma_{lg} S_{lg}(t) + \sigma_{sl} S_{sl}(t) + \sigma_{sg}  (S_{\infty} -S_{sl}(t) )
  \label{STE_GP}
\end{equation}
where $\sigma_{lg}$, $\sigma_{sl}$, and $\sigma_{sg} $ are the liquid-gas, solid-liquid, and solid-gas surface tension coefficients, respectively. The surface area of the drop at a given position is denoted as  $S_{lg}$. The solid-liquid surface area $S_{sl}=\pi x_{cl}^2$, where $x_{cl}$ is the x-coordinate of the contact line. The area of the solid surface is $S_{\infty}$, and it is considered $S_{\infty}\gg S_{sl}$. The gas-solid area is then $S_{\infty}- S_{sl}$. As shown in Fig. \ref{fig:Theoretical_DefDraw}, $S_{lg}$ and $S_{sl}$ will change over time as the contact line moves during drop oscillation. 

If we take the equilibrium state with zero-gravity as the reference (denoted by the subscript $0$), the surface energy can be rewritten as
\begin{align}
  E_s(t) &= E_s'(t) - E_{s,0}'= \sigma_{lg} (S_{lg}(t)-S_{lg,0} ) - \sigma_{lg} cos(\theta) \pi [(x_{cl}(t))^2 - x_{cl,0}^2]\,. 
 \label{eq:STE_wrt_Eq}
\end{align}
With the Young's Equation for contact angle $\theta$:
\begin{equation}
  \sigma_{sg}  = \sigma_{sl} + \sigma_{lg} cos \theta\,,
 \label{eq:YoungsEqn2}
\end{equation}
Eq.\ \eqr{STE_wrt_Eq} can be written as
\begin{equation}
  E_s(t) = \sigma_{lg} (S'(t) - S_0'),
 \label{eq:ES2}
\end{equation}
where $S'$ is the modified drop surface area, defined as
\begin{align}
  S'(t) & = S_{lg}(t) - (cos \theta) \pi [x_{cl}(t)]^2\,,
  \label{eq:modified_drop_surf_area}
\end{align}
which at the reference state is
\begin{align}
   S_0' &= S_{lg,0} - (cos \theta) \pi x_{cl,0}^2.
    \label{eq:ModSurfArea_equil}
\end{align}
where  $S_{lg,0} = 2 \pi R_0^2 (1- cos \theta)$ and $x_{cl,0} = R_0 sin \theta$ are the liquid-gas surface area and the $x$-coordinate of the contact line corresponding to the reference state. 

Following the previous study of Celestini and Koffman \cite{Celestini_2006a},  the drop shapes under first-mode oscillations are considered to be similar to the equilibrium shapes of the drop under different body forces (the body force acceleration is taken to be positive along the negative $z$-direction). The equilibrium drop theory in Appendix \ref{sec:equilibrium} is used to obtain the relation between $S'$ and the drop centroid $z$-coordinate, $z_c$, see Appendix \ref{sec:equilibrium_varyg}. For small-amplitude oscillations, the deviation from the equilibrium state is small, \ie, $|z_c-z_{c,0}|\ll R_0$. Therefore, we keep the two leading terms in the series expansion of $S'$ near the equilibrium state, 
\begin{equation}
 \frac{S'-S_0'}{S_0'} = \eta \left(\frac{z_c-z_{c,0}}{R_0}\right)^2 - \xi \left(\frac{z_c-z_{c,0}}{R_0}\right)^3 + O\left[\left(\frac{z_c-z_{c,0}}{R_0}\right)^4\right]\,. 
 \label{eq:FindingHTheta}
\end{equation}
Since the first order derivative is zero at  $z_c=z_{c,0}$, the linear term vanishes and the quadratic term is the leading-order term. The cubic term is kept here to account for the different behavior for $z_c<z_{c,0}$ (sessile drops) and $z_c>z_{c,0}$ (pendant drops).  The coefficients $\eta$ and $\xi$  for a given $\theta$ are determined based on the equilibrium drop results, see for example Fig.\ \ref{fig:equil_shape_g} for $\theta=\ang{90}$ and $\ang{130}$ in Appendix \ref{sec:equilibrium_varyg}. 

Finally, substituting Eq.\ \eqr{FindingHTheta} into Eq.\ \eqr{ES2}, the surface energy $E_s$ can be expressed as 
\begin{equation}
	E_s = \sigma_{lg} S_0' \left[\eta \left(\frac{z_c-z_{c,0}}{R_0}\right)^2 - \xi \left(\frac{z_c-z_{c,0}}{R_0}\right)^3\right]\, .
 \label{eq:ES3}
\end{equation}

\begin{figure}
 \centering
 \includegraphics[width=0.5\textwidth]{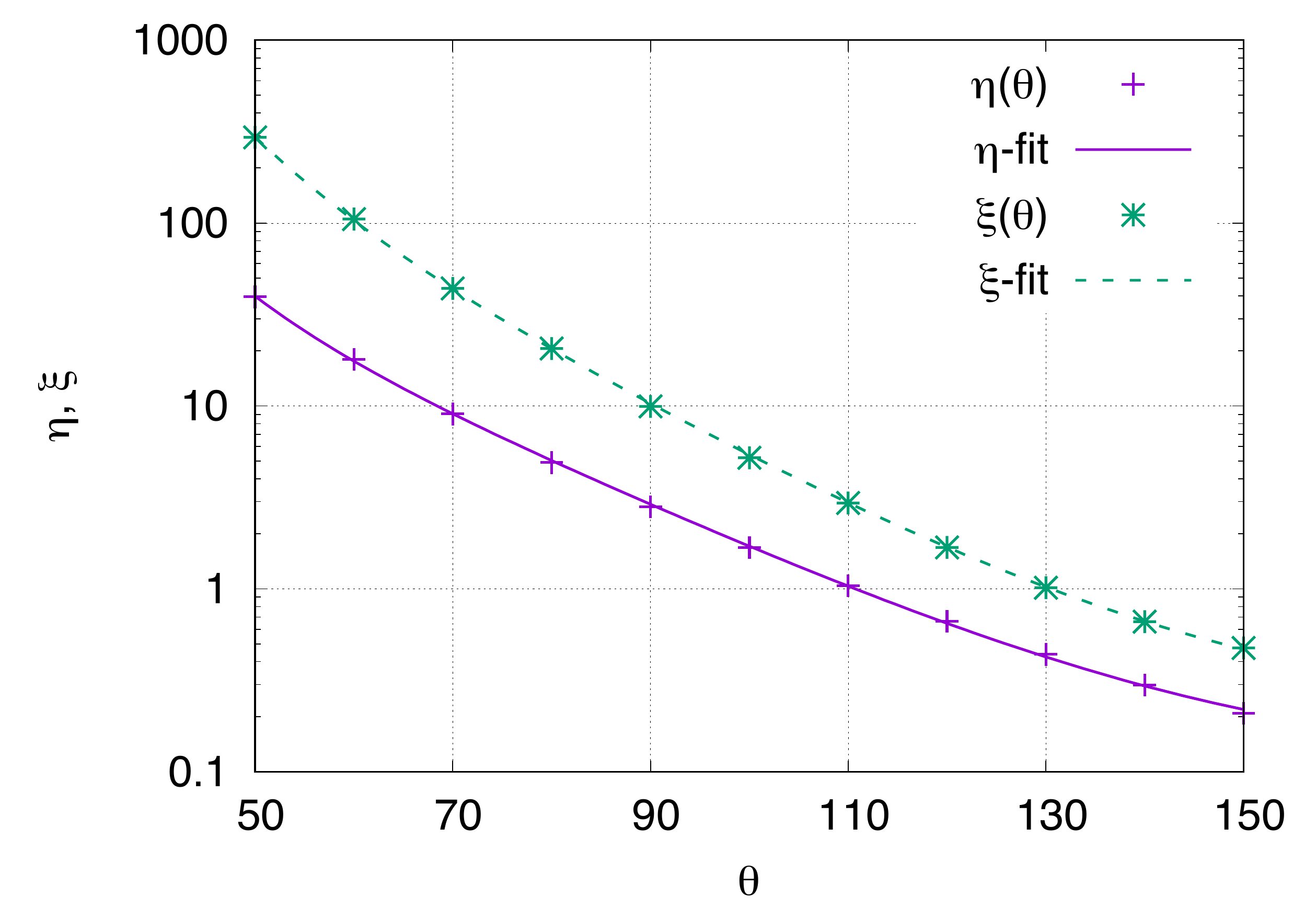}
 \caption{Variation of model parameters $\eta$ and $\xi$ with $\theta$ (degree). }
 \label{fig:model}
\end{figure}

\subsubsection{Potential energy}
For finite gravity $g$, the gravitational potential energy of the drop with respect to the reference state is 
\begin{equation}
  E_g(t) = m_d g (z_c-z_{c,0})\, ,
 \label{eq:Eg}
\end{equation}
where $m_d=\rho_l V_d$ is the mass of the drop. The overall potential energy of the sessile drop $E_p$ is the sum of $E_s$ and $E_g$, namely 
\begin{equation}
 	E_p = E_s + E_g \,. 
 \label{eq:E_p}
\end{equation}
The equilibrium state varies with $g$ and is located at the minimum of $E_p$. 
The centroid $z$-coordinate at the equilibrium state for finite gravity, $z_{c,1}$, can be determined from Eqs.\ \eqr{ES3} and \eqr{Eg} as
\begin{equation}
  \od{E_p }{z_c}\rvert_{z_c=z_{c,1}} = - \frac{3\xi \sigma_{lg} S_0'}{R_0^3} (z_c-z_{c,0})^2+\frac{2\eta \sigma_{lg} S_0'}{R_0^2} (z_c-z_{c,0}) + m_d g = 0\,. 
 \label{eq:E_g_Es_min}
\end{equation}
The quadratic equation can be solved and the relevant root is given as 
\begin{equation}
 	\frac{z_{c,1}-z_{c,0}}{R_0} = \frac{\eta - \sqrt{\eta^2 + \xi Bo_0}}{3\xi }\,.
 \label{eq:zc1}
\end{equation}

The potential energy $E_p$ can be rearranged to a series expansion near $z_c=z_{c,1}$ as
\begin{equation}
	E_p- E_{p,z_{c,1}}= \sigma_{lg} S_0' \left[\eta' \left(\frac{z_c-z_{c,1}}{R_0}\right)^2 - \xi' \left(\frac{z_c-z_{c,1}}{R_0}\right)^3\right]\, ,
 \label{eq:ES_Eg_2}
\end{equation}
where $\xi' = \xi$ and 
\begin{equation}
	\eta' = \eta \sqrt{1+\frac{\xi Bo_0}{\eta^2}} = \eta \sqrt{1+\frac{\hat{\xi} Bo}{\eta^2}}\,,  
 \label{eq:ES_Eg_2}
\end{equation}
 where $\hat{\xi} = \xi R_0^2/R_d^2$, which also depends only on $\theta$ (see Eq.\ \eqr{R0}). 
For small-amplitude oscillations near the equilibrium state, \ie, $|z_c-z_{c,1}|\ll R_0$, the cubic term can be dropped as well, yielding
\begin{equation}
	E_p =  E_{p,z_{c,1}} + \sigma_{lg} S_0' \eta' \left(\frac{z_c-z_{c,1}}{R_0}\right)^2 \, .
 \label{eq:ES_Eg_3}
\end{equation}

\subsubsection{Kinetic energy}
\begin{figure}
 \centering
 \includegraphics[width=1\textwidth]{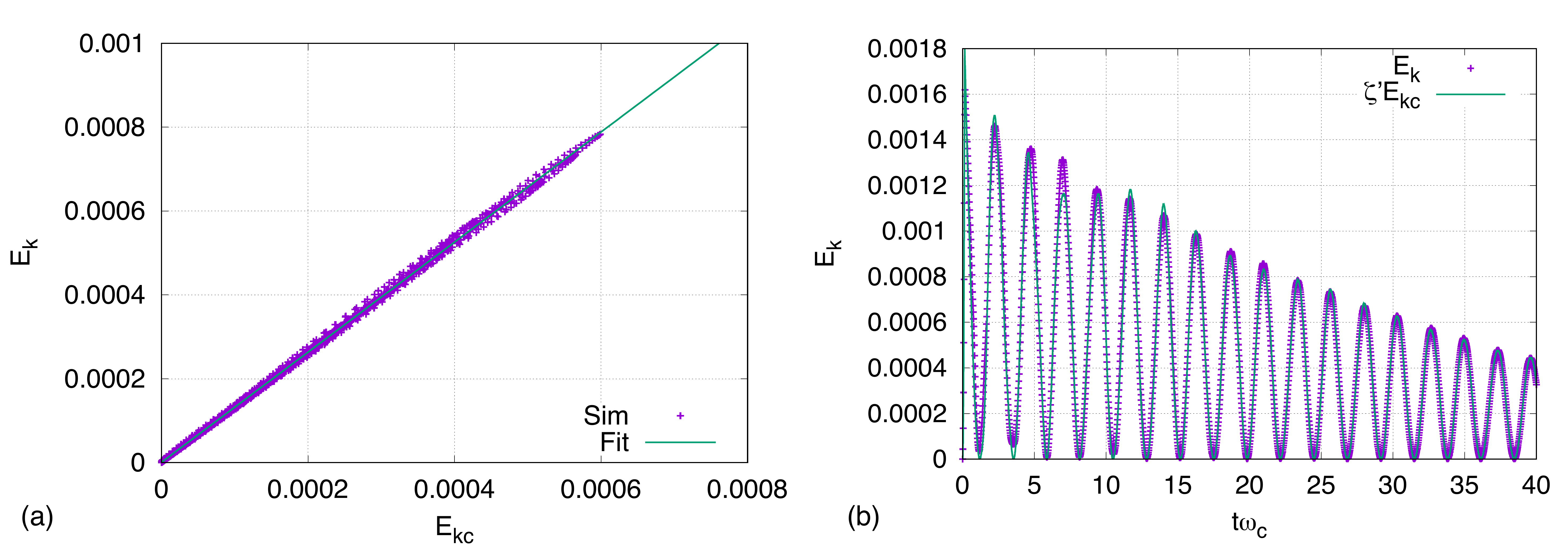}
 \caption{(a) $E_k$ as a function of $E_{kc}$ and (b) the temporal evolutions for $E_k$ and $\zeta E_{kc}$ for $\theta=\ang{130}$ and $Bo=0$.  }
 \label{fig:KE_drop_130}
\end{figure}

The kinetic energy of the sessile drop, expressed as
\begin{equation}
	E_k = \rho_l\int_{V_d} \frac{ |\bs{u}|^2}{2} dV \, ,
 \label{eq:EK1}
\end{equation}
varies over time as the drop oscillates. 
If the drop is treated as a rigid object moving with the centroid velocity $u_c$, the kinetic energy is 
\begin{equation} 
	E_{k,c}= \frac{m_d |\bs{u_c}|^2}{2} =  \frac{m_d w_c^2}{2} =  \frac{m_d}{2} \left(\od{z_c}{t}\right)^2\, .
\end{equation}
where $w_c$ is the $z$-component of centroid velocity, $\bs{u_c}$.
Due to the internal flows induced by the shape oscillations, the kinetic energy $E_k$ is generally larger than $E_{k,c}$. The simulation results of $E_k$ and $E_{kc}$ for $\theta=130$\textdegree\ and $Bo = 0$ are shown in Fig.\ \ref{fig:KE_drop_130}. It is shown that $E_k$ is approximately a linear function of $E_{kc}$ for small-amplitude oscillations. This conclusion has been verified for other $\theta$ not plotted here. This important observation provides a convenient way to approximate $E_k$, 
\begin{equation}
	E_k \approx \zeta' E_{k,c}=  \frac{  \zeta' m_d}{2} \left(\od{z_c}{t}\right)^2\, ,
 \label{eq:EK2}
\end{equation}
where $\zeta'$ is the kinetic energy correction factor. For the small-amplitude first-mode oscillations, $\zeta'$ varies little over time and depends only on $\theta$  and $Bo$, \ie, $\zeta'=\zeta'(\theta,Bo)$. For a given combination of $\theta$ and $Bo$, $\zeta'$ can be obtained by fitting the corresponding simulation results, see for example Fig.\ \ref{fig:KE_drop_130}(a) for $\theta=\ang{130}$ and $Bo=0$. The temporal evolutions of $E_k$ and $\zeta' E_{k,c}$ with the fitted value of $\zeta'$ are plotted in Fig.\ \ref{fig:KE_drop_130}(b). It is seen that $\zeta' E_{k,c}$ agrees well with $E_k$ for all time. The small discrepancies at early time are due to the high-order modes triggered. 

\begin{figure}
 \centering
 \includegraphics[width=1\textwidth]{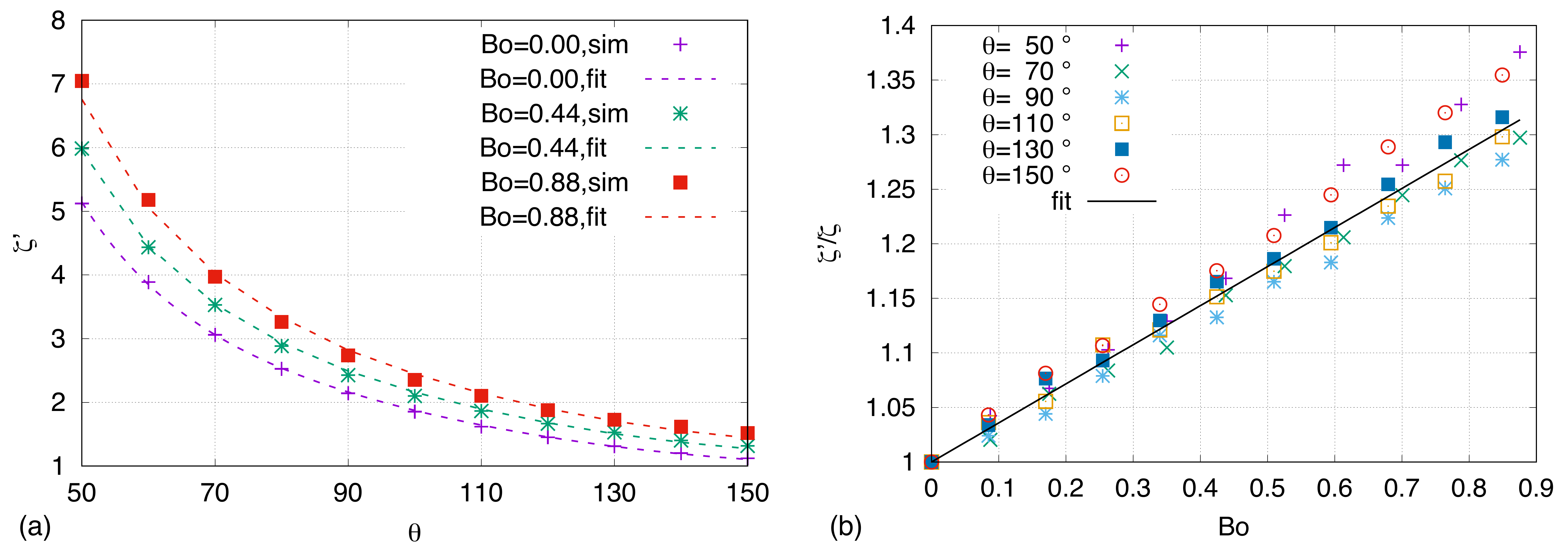}
 \caption{(a) Kinetic energy correction factor $\zeta'$ as a function of different $\theta$ (degree) and $Bo$. The symbols are simulation results and the dashed lines are approximations made by Eqs.\ \eqr{zeta_model} and \eqr{zeta_fit}. (b)  $\zeta'/\zeta$ as a function of $Bo$ for different $\theta$. The black solid line is the linear fit of  the results of all the cases.  }
 \label{fig:zeta}
\end{figure}

The values of $\zeta'$ for different $\theta$ and $Bo$ are shown in Fig.\ \ref{fig:zeta}(a). It is observed that $\zeta'$ generally decreases with $\theta$. When  $Bo \to 0$ and $\theta\to \ang{180}$, the sessile drop reduces to a free drop, the shape oscillations vanish and $\zeta'=1$. When $\theta$ decreases, the equilibrium shape of the sessile drop is more and more deviated from the spherical shape, as a consequence, the ratio between $E_{k}$ and $E_{kc}$ increases. Furthermore, for a given $\theta$, it is shown in Fig.\ \ref{fig:zeta}(b) that $\zeta'$ approximately increases linearly with $Bo$. Though $\zeta'$ varies with both $\theta$ and $Bo$, if $\zeta'(\theta,Bo)$ is normalized by $\zeta(\theta)=\zeta(\theta, Bo=0)$, the results for different $\theta$ tend to collapse a similarity solution. Therefore, $\zeta'$ can be expressed as
\begin{equation}
	\zeta'(\theta,Bo) = \zeta(\theta) (1+ \alpha Bo)\,, 
	\label{eq:zeta_model}
\end{equation}
where the coefficient $\alpha$ is a constant.

\subsubsection{Frequency of the first-mode oscillation}
Based on the inviscid flow assumption, the total energy is constant. With Eqs.\ \eqr{ES_Eg_3} and \eqr{EK2}, it yields 
\begin{equation}
 \frac{dE_{tot}}{dt} = \frac{\sigma_{lg}  S_0' \eta' }{R_0^2} 2 (z_c - z_{c,1}) \frac{d z_c}{dt} + \zeta m_d \frac{dz_c}{dt} \frac{d^2z_c}{dt^2} =0\, ,
 \label{eq:dEKdt}
\end{equation}
which can be simplified to the form for a harmonic oscillator, 
\begin{equation}
	 k (z_c - z_{c,1}) + m_d \frac{d^2  (z_c - z_{c,1})}{dt^2} = 0,
 \label{eq:HarmonicOscillator}
\end{equation}
where $k = {2 \sigma_{lg} \eta' S_0'}/{(\zeta R_0^2)}$. The  angular frequency of the harmonic oscillator is $\omega=\sqrt{k/m_d}$. Combining Eqs.\ \eqr{R0} and \eqr{ModSurfArea_equil}, the first-mode frequency can be obtained
\begin{equation}
 \frac{\omega}{\omega_c} 
 = \sqrt{\frac{6 \eta'}{\zeta'}} 
 \label{eq:NormalizedFreq2}
\end{equation}
With Eqs.\ \eqr{ES_Eg_2} and \eqr{zeta_model}, the frequency can be also written as 
\begin{equation}
 \frac{\omega}{\omega_c} 
	= \sqrt{\frac{6 \eta}{\zeta}} f(\theta, Bo),
 \label{eq:NormalizedFreq3}
\end{equation}
where $\omega/\omega_c|_{Bo=0} = \sqrt{6\eta/\zeta}$ is the expression for the frequency in the zero-$Bo$ limit, and $f$ is the correction function for finite $Bo$,
\begin{equation}
	f(\theta,Bo)	= 	\frac{\left(1+\frac{\hat{\xi}}{\eta^2} Bo \right)^{1/4}}{(1+\alpha Bo)^{1/2}}\, .
 \label{eq:correction_finiteBo}
\end{equation}
The parameters $\eta$, $\hat{\xi}$, and $\zeta$ depend on $\theta$ only and $\alpha$ is a constant. With $\eta$ and $\hat{\xi}$ determined by the equilibrium sessile drop theory and $\zeta$ and $\alpha$ by the simulation results, Eqs.\ \eqr{NormalizedFreq2}-\eqr{correction_finiteBo} can be used to predict the first-mode oscillation frequency for the sessile drop with FCL for different contact angles and Bond numbers. Due to the truncation of higher-order terms in Eq.\ \eqr{FindingHTheta}, Eq.\ \eqr{correction_finiteBo} is strictly valid for small $(z_c-z_{c,1})/R_0$ or for small $Bo$.

\subsubsection{Fitted correlations for $\eta$, $\hat{\xi}$, and $\zeta$}
To make Eq.\ \eqr{NormalizedFreq2}-\eqr{correction_finiteBo} easier to use, explicit expressions for $\eta$, ${\xi}$, and $\zeta$ are  obtained by fitting the results for the equilibrium sessile drop theory and the simulations. The same functional form is applied to all three parameters, 
\begin{align}
	\log(\eta(\theta)) = c_0+c_1 (1+\cos\theta) +\left[\exp\left(\frac{(1+\cos\theta) ^{c_2} }{c_3} \right)-1\right]\,,
	\label{eq:eta_fit}\\
	\log({\xi}(\theta)) = e_0+e_1 (1+\cos\theta) +\left[\exp\left(\frac{(1+\cos\theta) ^{e_2} }{e_3} \right)-1\right]\,,
	\label{eq:xi_fit}\\
	\log(\zeta(\theta)) = b_0+b_1 (1+\cos\theta) +\left[\exp\left(\frac{(1+\cos\theta) ^{b_2} }{b_3} \right)-1\right]\,. 
	\label{eq:zeta_fit}
\end{align}
The fitted constants are $[c_1, c_2, c_3, c_4]=[-1.92, 2.97, 7.01, 59.4]$, $[e_1, e_2, e_3, e_4]=[-1.26,3.55,5.32,18.5]$, $[b_0,b_1,b_2,b_3] = [0, 0.753, 5.87, 54.6]$ for $\eta$, ${\xi}$, and $\zeta$, respectively. Substituting Eq.\ \eqr{zeta_fit} to Eq.\ \eqr{zeta_model}, the coefficient $\alpha$ can be determined by fitting the results of $\zeta$ for all the cases considered and it is found that $\alpha\approx 0.358$. 

The fitting functions Eqs.\ \eqr{eta_fit}-\eqr{xi_fit} are plotted in Fig.\ \ref{fig:model} and are found to yield very  good approximations for the exact results of $\eta$ and $\xi$.  For large $\theta$, $1+\cos(\theta)$ is small, then the expression reduces to a linear function, \eg, $\log \eta \approx c_0+c_1 (1+\cos\theta) $. The linear relations between $\log (\eta), 
\log(\xi), \log(\zeta)$ and $(1+\cos\theta) $ actually hold for all hydrophobic cases $\theta>\ang{90}$. The term $[\exp\left({(1+\cos\theta) ^{c_2} }/{c_3} \right)-1]$ is mainly used to account for the deviation of the hydrophilic cases from the linear function. 

The approximations of $\zeta'$ (Eqs.\ \eqr{zeta_model} and \eqr{zeta_fit}) are compared with the simulation results in Fig.\ \ref{fig:zeta}(a). Again, a good agreement is achieved. In the limit of $\theta  \to \ang{180}$ $\zeta=1$, so $b_0=0$. Since the above fitting functions and coefficients (Eqs.\ \eqr{zeta_model}, \eqr{eta_fit}-\eqr{zeta_fit}) are obtained from the data for $\ang{50}\le \theta \le \ang{150}$, caution is required if they are used for parameters outside of this range. 

\section{Results}
\label{sec:results}

\subsection{First oscillation mode}
\label{sec:1st_mode}

\begin{figure}
 \centering{\includegraphics[width=0.99\textwidth]{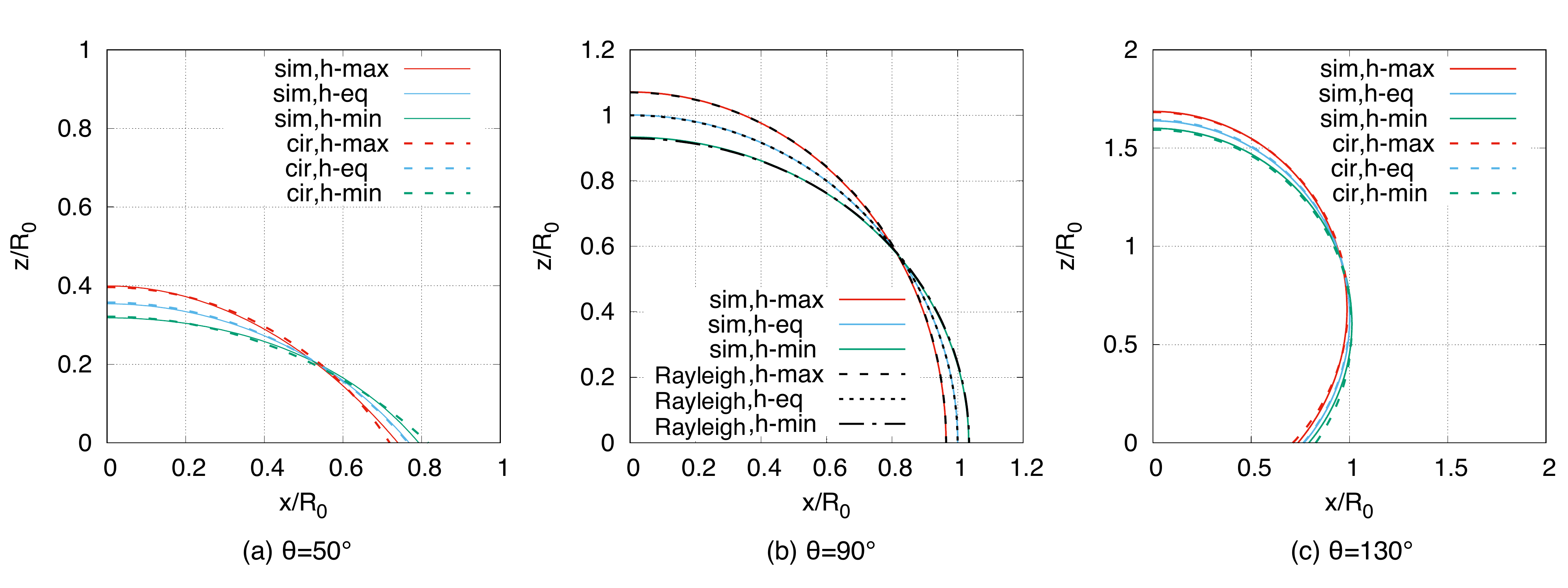}}
 \caption{Simulation results for the drop oscillations induced by IC1 for zero gravity and different contact angles: (a) $\theta=\ang{50}$, (b) $\theta=\ang{90}$, and (c) $\theta=\ang{130}$. Different colors indicate the  representative drop shapes corresponding to the maximum, equilibrium, and minimum drop height in one first-mode oscillation cycle. The solid lines represent simulation results, while the dashed circles with the same color in (a) and (c) represent the spherical caps with different centroid locations and the same volume.  The black dashed, long-dashed, and dash-dot lines in (b) represent the second Rayleigh mode for a free drop with mode amplitude corresponding to the simulation results.  }
 \label{fig:modeIC1_sim}
\end{figure}

The simulation results for the drop oscillations induced by IC1 for zero gravity and different contact angles ($\theta=\ang{50}$, $\ang{90}$, and $\ang{130}$) are shown in Fig.\ \ref{fig:modeIC1_sim}. The instantaneous drop shapes correspond to the maximum, equilibrium, and minimum drop heights ($h_{max}$, $h_{eq}$, $h_{min}$), respectively. Since gravity is zero, the equilibrium drop surfaces are circular arcs {(spherical caps in 3D)}.  The first-mode shape oscillation is induced by the translation of centroid. When the drop centroid is above the equilibrium position, the arc radius must be reduced due to mass conservation, or vice versa. If the solid surface allows an arbitrary contact angle, the drop shapes would remain as circular arcs of different radii. However, as the contact angle is fixed for FCL condition, the exact shape of the drop deviates from the corresponding circular arc and is adjusted for the specified contact angle. It can be observed {from Figs.\ \ref{fig:modeIC1_sim}(a) and (c)} that, the drop contours away from the equilibrium state ($h_{max}$ and $h_{min}$) agree with the corresponding circular arcs in general, except in small regions near the solid surface.

The case for $\theta=\ang{90}$ is a special case. With the slip wall boundary condition, the bottom wall for $\theta=\ang{90}$ is identical to a symmetric boundary. Therefore, the sessile drop with first-mode oscillation will be identical to the top half of a free drop with twice the volume undergoing a second-mode oscillation \cite{Lyubimov_2006a, Bostwick_2014a}. The drop shapes in Fig.\ \ref{fig:modeIC1_sim}(b) are expected to agree well the second harmonic modes described by the Legendre polynomials, see Eq.\ \eqr{Ledendre} and that is exactly what is observed in Fig.\ \ref{fig:modeIC1_sim}(b). 

\subsubsection{Effect of contact angle on oscillation frequency}
\label{sec:effect_contact_angle}

\begin{figure}
 \centering{\includegraphics[width=1\textwidth]{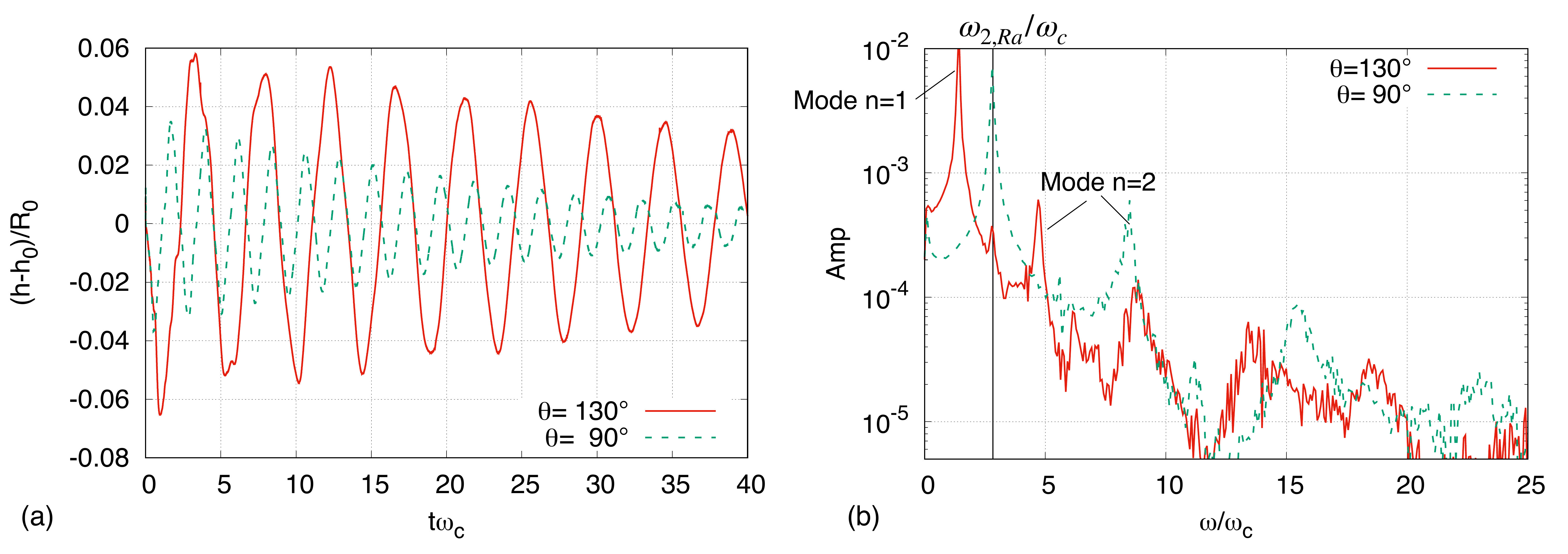}}
 \caption{(a) Temporal evolution and (b) frequency spectrum of the drop height for $\theta = \ang{130}$ and $\ang{90}$, $Bo = 0$, and $Oh = 0.00571$. The IC1 approach is used to initiate the drop shape oscillation, which primarily activates the first mode. }
 \label{fig:modeIC1_theta}
\end{figure}

The temporal evolution of the drop height $h$ (see Fig.\ \ref{fig:schematic}) is measured in the simulation. The results for $\theta=\ang{90}$ and 130\textdegree\  are shown in Fig.\ \ref{fig:modeIC1_theta}(a). Fourier transforms are performed to obtain the frequency spectra, see Fig.\ \ref{fig:modeIC1_theta}(b). Though other high-order modes are also triggered by the IC1 method, but their amplitudes are small and decay much faster than the first mode. The first mode clearly dominates in the spectrum. It is further observed that, the first-mode frequency increases when the contact angle decreases from $\ang{130}$ to 90\textdegree. The trend for the frequency variation over $\theta$ is consistent with former studies for sessile drops with PCL \cite{Strani_1984a, Bostwick_2014a}. The first-mode frequency for $\theta=\ang{90}$ agrees well the normalized Rayleigh frequency for the second mode ($\omega_{2,Ra}/\omega_c$), since the first mode of the sessile drop with $\theta=\ang{90}$ is similar to the second mode for a free drop, as explained above. 

\begin{figure}
 \centering
 \includegraphics[width=0.5\textwidth]{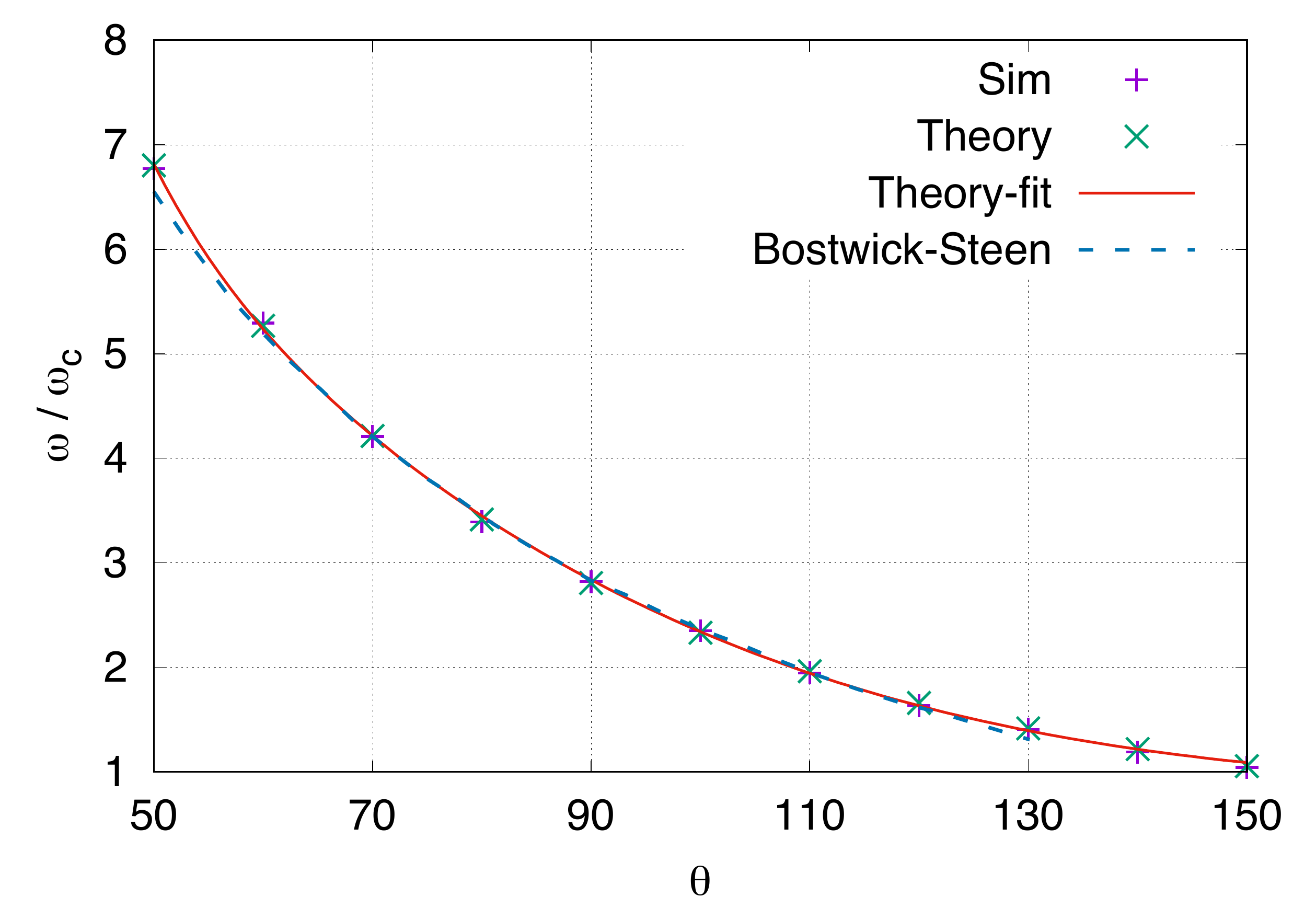}
 \caption{The theoretical predictions (Eq.\ \eqr{NormalizedFreq2}) and simulation results for the first mode frequency for different $\theta$ (degree), compared to the theory of Bostwick-Steen \cite{Bostwick_2014a}. The theoretical results using the exact values of $\eta$, $\hat{\xi}$, and $\zeta$ are represented by symbols, while those using the fitted functions (Eqs.\ \eqr{eta_fit},\eqr{xi_fit}, and \eqr{zeta_fit}) by the solid line. }
 \label{fig:freq_mode1_Bo0}
\end{figure}

The  first-mode oscillation frequencies $\omega/\omega_c$ {predicted by the theoretical model described in section \ref{sec:theory}} are compared with the simulation results in Fig.\ \ref{fig:freq_mode1_Bo0}. The model predictions agree remarkably well with the simulation results for the whole range of $\theta$ considered.  If the fitted expressions of $\eta$, $\xi$, and $\zeta$ (Eqs.\ \eqr{eta_fit},\eqr{xi_fit}, and \eqr{zeta_fit}) are used, then Eq.\ \eqr{NormalizedFreq2} turns into a fully explicit expression for the first-mode frequency for the sessile drop with FCL for arbitrary contact angles and Bond numbers. The theoretical predictions using the fitted functions match perfectly with the theoretical predictions with exact values of $\eta$, ${\xi}$, and $\zeta$, see Fig.\ \ref{fig:freq_mode1_Bo0}. 

{
The frequencies predicted by the inviscid model of Bostwick and Steen \cite{Bostwick_2014a} are also shown for comparison. It is seen that the Bostwick-Steen model predictions agree well with the present simulation results in general, though the Bostwick-Steen theory slightly underestimates $\omega/\omega_c$ for $\theta<\ang{60}$ and $\theta<\ang{120}$. The discrepancy between Bostwick-Steen model and the present results are probably attributed to the contact-line velocity boundary condition (BC) used in their model. The BC is strictly valid only when $\theta=\ang{90}$, see Appendix~\ref{sec:BC_CL}. That is why the frequencies predicted by their model agree very well with the present results for $\theta=\ang{90}$ and some small discrepancies arise for large $|\theta - \ang{90}| $.  }

\subsubsection{Effect of Bond number on oscillation frequency}
\label{sec:effct_Bo}
%
%

\begin{figure}
 \centering
 \includegraphics[width=1\textwidth]{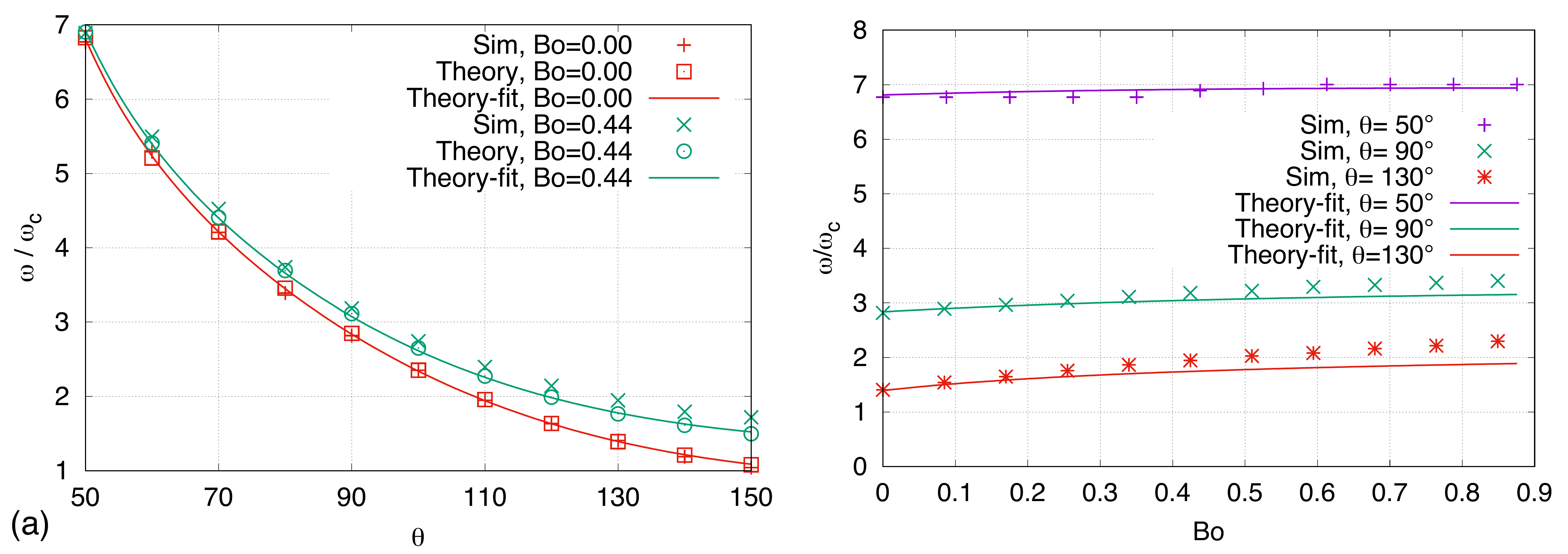}
 \caption{First mode oscillation frequency as a function of $\theta$ (degree) and $Bo$. The results for simulations and theoretical model are denoted by symbols, while the theoretical model predictions with fitting functions are denoted by solid lines. }
 \label{fig:freq_mode1}
\end{figure}

The simulation results for the first-mode frequencies for different contact angles and Bond numbers are shown in Fig.\ \ref{fig:freq_mode1}. It is observed that $\omega/\omega_c$ increases with $Bo$ for all $\theta$. The increase of $\omega/\omega_c$ over $Bo$ is more profound for large $\theta$. The theoretical results are also plotted here for comparison, which agree very well with the simulations for small $Bo$, affirming that the model successfully captures the effect of Bond number. The model predictions using the fitting functions are also plotted, which again agree well with the predictions using the exact $\eta'$ and $\zeta'$. Because in the model, the high-order terms were truncated in the expansion of surface area (see Eq.\ \eqr{ES3}), the theoretical model is strictly valid only for small $Bo$, that is why the model predictions deviate from the simulation results for large $Bo$. The deviation is more significant for large $\theta$. Nevertheless, for the range of parameters considered, the model predictions are still very good approximations for the first-mode frequencies. The maximum error of the model compared to the simulation results (\ie, for the case $Bo=0.88$ and $\theta=\ang{150}$) is about 12\%. 

The theoretical model, as formulated in Eq.\ \eqr{NormalizedFreq2}, is also useful in revealing the physics behind the variation of the first-mode frequency with $\theta$ and $Bo$. First of all, as shown in Fig.\ \ref{fig:freq_mode1}(a), $\omega/\omega_c$ decreases with $\theta$ for all $Bo$; this is mainly due to the decrease of $\eta$ over $\theta$. As shown in Eq.\ \eqr{FindingHTheta}, $\eta$ measures the concavity of the surface area (or surface energy) as a function of the deviation of the drop centroid  at the equilibrium state. When $\theta$ increases, the increase of surface area, $S'-S_0'$, for a given deviation of the centroid location, $z_c-z_{c,0}$, is reduced and thus $\eta$ decreases. For $\theta \rightarrow \ang{180}$, $S'-S_0'$ will be identical to zero. As a result, $\eta = 0$ and $\omega = 0$. As can be seen in Fig. \ref{fig:equil_shape_g}, as $\theta$ decreases from $\ang{130}$ to $\ang{90}$, the $S'$-$z_c$ profiles become more concave up. Correspondingly, $\eta$ increases from about 0.21 to about 2.81, see Fig. \ref{fig:model}.

Furthermore, it is also shown that the frequency increases with $Bo$  for all $\theta$. When gravity is present, the equilibrium centroid location, $z_{c,1}$, is reduced, compared to that without gravity, $z_{c,0}$. The expansion of surface area at the corresponding equilibrium state leads leads to a correction factor for $\eta$ and the effective $\eta'$ becomes $\eta\sqrt{1+\hat{\xi} Bo/\eta^2}$. The coefficient of the cubic term in Eq.\ \eqr{FindingHTheta}, $\hat{\xi}$, is due to the asymmetry between the left (sessile drops, $dz_c<0$) and right (pendant drops, $dz_c>0$) branches of the $S'$-$z_c$ profiles, see Fig.\ \ref{fig:equil_shape_g}. For a given $z_c-z_{c,0}$, $S'-S_0'$ is larger for sessile drops than for pendant drops. As a result, $\hat{\xi}$ is always positive for all $\theta$. Therefore, the correction function, $\eta'$, monotonically increases with $Bo$. The correction function $f$ (Eq.\ \eqr{correction_finiteBo}) can be further approximated for small $Bo$ ($|\alpha Bo| \ll 1$ and $|\hat{\xi}Bo/\eta^2|\ll 1$) using Taylor expansion keeping only the first-order term, 
\begin{equation}
	f \approx
	1+(\frac{\hat{\xi}}{4\eta^2} - \frac{\alpha}{2}) Bo\,.
\end{equation}
It can be shown that the coefficient for $Bo$, $(\frac{\hat{\xi}}{4\eta^2} - \frac{\alpha}{2})>0$ for all $\theta$, as a result, both $f$ and $\omega/\omega_c$ increase with $Bo$. 

Finally, the increase of $\omega/\omega_c$ over $Bo$ becomes more profound as $\theta$ increases. This is because the coefficient of $Bo$, \ie, $\hat{\xi}/\eta^2$, increases with $\theta$. For $\theta=\ang{50}$, the change of $\omega/\omega_c$ over $Bo$ is very small, and thus is hard to capture by the simulation. This is the reason for the noises in the simulation results for $\theta=\ang{50}$ shown in Fig.\ \ref{fig:freq_mode1}(b). The resolution of the frequency spectra depends on the simulation time. The simulation would need to be run for a much longer time to accurately capture the small frequency change. Therefore, the theoretical model is in particular useful in predicting the oscillation frequency for small $\theta$.

\subsection{High-order oscillation modes}
\label{sec:higher_order_mode}
\begin{figure}
 \centering{\includegraphics[width=1\textwidth]{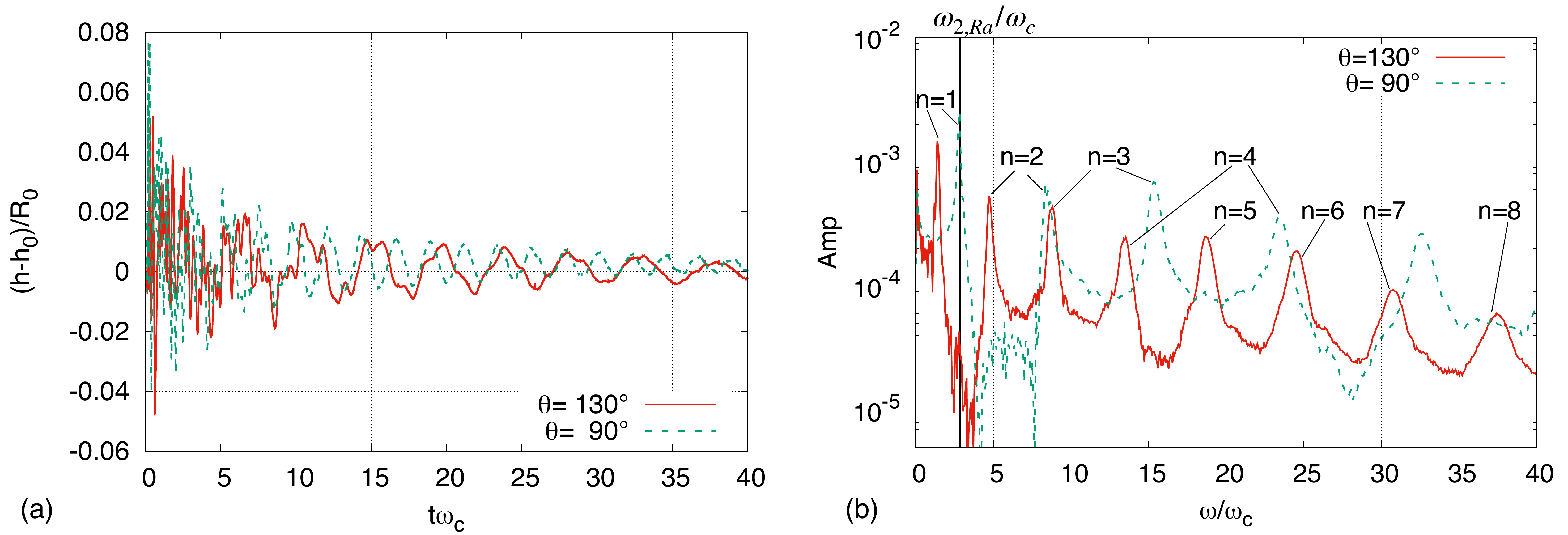}}
 \caption{(a) Temporal evolution and (b) frequency spectrum for the drop height for $\theta = \ang{90}$ and $\ang{130}$, $Bo = 0$, and $Oh = 0.00571$. The IC2 approach is used to initiate the drop shape oscillation, which activates the first and also the high-order modes.}
 \label{fig:modeIC2_theta}
\end{figure}

As the damping rates for high-order ($n>1$) modes are higher than that for the first mode, even though the high-order modes are triggered by IC1, the energy contained in the modes is low, making them hard to be identified in the frequency spectra, see Fig.\ \ref{fig:modeIC1_theta}(b). To accurately measure the frequencies of the high-order modes, the IC2 method is used to initiate drop oscillations. For IC2, the initial energy distribution is biased toward the high-order modes. The temporal evolutions and frequency spectra for the drop height for $\ang{90}$ and $\ang{130}$ are shown in Fig.\ \ref{fig:modeIC2_theta}. It can be seen that the amplitudes of the high-order modes in the spectra are higher than those in Fig.\ \ref{fig:modeIC1_theta}, though the frequencies for all the modes induced by IC2 generally agree well with those induced by IC1. For example, the first-mode frequency for $\theta=\ang{90}$ also matches the second-mode Rayleigh frequency for the corresponding free drop ($\omega_{2,Ra}/\omega_c$). 

\begin{figure} 
 \centering{\includegraphics[width=1\textwidth]{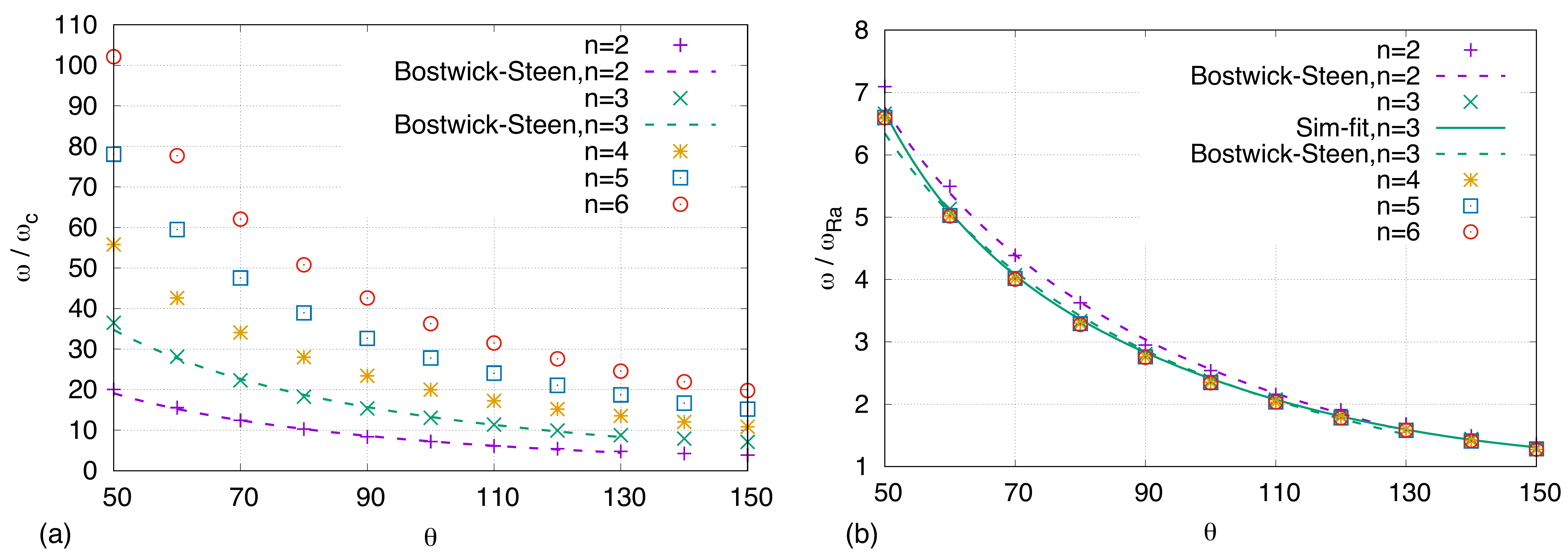}}
 \caption{Variation of the oscillation frequencies for modes $n=2$ to 6 over the contact angle $\theta$ (degree), for $Bo = 0$, $Oh = 0.00571$. (a) $\omega$ is normalized by $\omega_c$; (b) $\omega$ is normalized by the Rayleigh frequency $\omega_{Ra}$. }
 \label{fig:freq_mode2-6_theta}
\end{figure}

The frequencies of the high-order modes for different $\theta$ are shown in Fig.\ \ref{fig:freq_mode2-6_theta}. It is observed that the frequencies for all the modes presented here ($n=2$ to 6) decrease with $\theta$, similar to the first mode. The Bostwick-Steen theory for $n=2$ and $n=3$ are also plotted, {which again agree well with the simulation results in general, except for very small and very large $\theta$, similar to the $n=1$ mode}. Furthermore, the oscillation frequencies $\omega_n/\omega_c$ for all $\theta$ increase with the mode number $n$. However, it is interesting to observe in Fig.\ \ref{fig:modeIC2_theta}(b) that the oscillation frequencies $\omega_n$ tend to collapse if they are normalized by the corresponding Rayleigh frequencies $\omega_{n,Ra}$. As $\theta$ increases, all the scaled frequencies, $\omega_n/\omega_{n,Ra}$, approach unity for all $n$. This is in agreement with the asymptotic limit for $\theta \to \ang{180}$, where the sessile drop becomes a free drop and the frequencies for $n\ge 2$ modes reduce to the corresponding Rayleigh frequencies. The profiles of $\omega/\omega_{Ra}$ for $n\ge 3$ modes agree very well for all $\theta$, while the frequencies for the second mode scale better for the hydrophobic ($\theta> \ang{90}$) than the hydrophilic cases ($\theta< \ang{90}$). This is probably because the equilibrium shape of a sessile drop for small $\theta$ becomes more like a lens and less similar to a sphere, which is assumed in Rayleigh's theory. The scaling relation indicates that $\omega_n \sim \sqrt{(n-1)n(n+2)}$, which approaches to $n^{3/2}$ as the mode number $n\to \infty$. Similar observations have been made for sessile drops with PCL \cite{Vukasinovic_2007a}.

\begin{figure} 
 \centering{\includegraphics[width=1\textwidth]{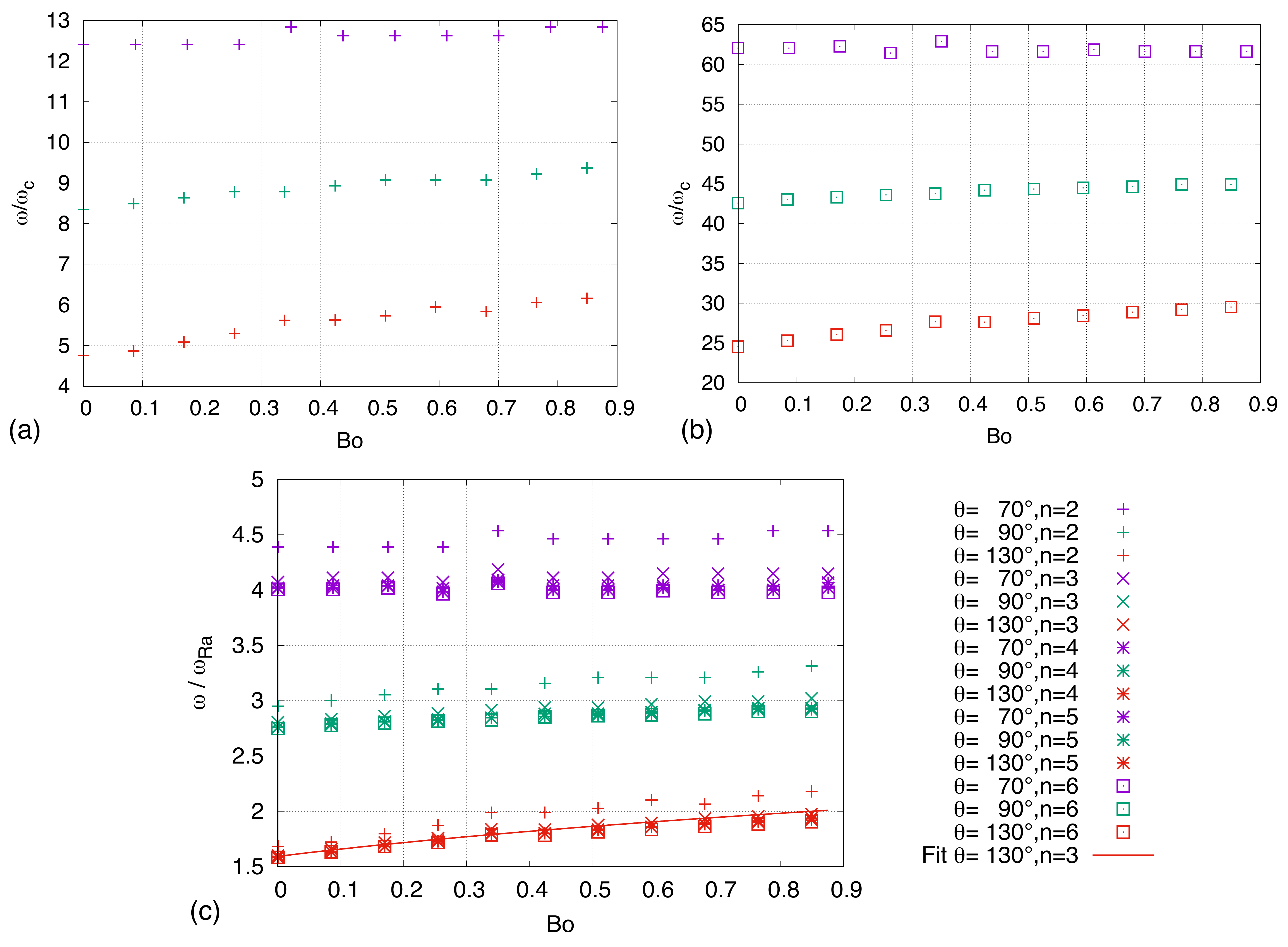}}
 \caption{Variation of the oscillation frequencies normalized by the capillary frequency, $\omega/\omega_c$, over $Bo$ for different contact angles $\theta=\ang{70},\ \ang{90},$ and $\ang{130}$ for modes (a) $n=2$ and (b) $n=6$. (c) Variation of oscillation frequencies normalized by the corresponding Rayleigh frequencies over $Bo$ for . }
 \label{fig:freq_mode2-6_Bo}
\end{figure}

The effect of $Bo$ on the high-order modes are shown in Fig.\ \ref{fig:freq_mode2-6_Bo}. Similar to the first mode, the frequencies of the high-order modes $\omega_n/\omega_c$ generally increase with $Bo$. The increase over $Bo$ is more gradual as $\theta$ decreases. For $\theta \le \ang{70}$, the variation of $\omega_n$ is small, which is hard to resolve via the simulation (due to relatively short simulation time and the resulting limited resolution in the frequency spectra). That is why the results for $\theta=\ang{70}$ show small fluctuations in Figs.\ \ref{fig:freq_mode2-6_Bo}(a) and (b). 

The scaling relation between $\omega$ and $\omega_{Ra}$ generally holds for finite $Bo$, see Fig.\ \ref{fig:freq_mode2-6_Bo}(c). Similar to the results for $Bo=0$, the scaled frequencies $\omega_n/\omega_{n,Ra}$ agree better for $n\ge 3$ than the second mode. The agreement is also better for large contact ($\theta=\ang{130}$) than small ($\theta=\ang{70}$) contact angles. Furthermore, the scaling relation between $\omega$ and $\omega_{Ra}$ becomes less accurate as $Bo$ increases. This is again probably related to the equilibrium shape of the sessile drop. For large $Bo$, the equilibrium drop in the limit of $\theta=\ang{180}$ will become more flattened and be deviated from the spherical shape. 

\begin{figure} 
 \centering{\includegraphics[width=1\textwidth]{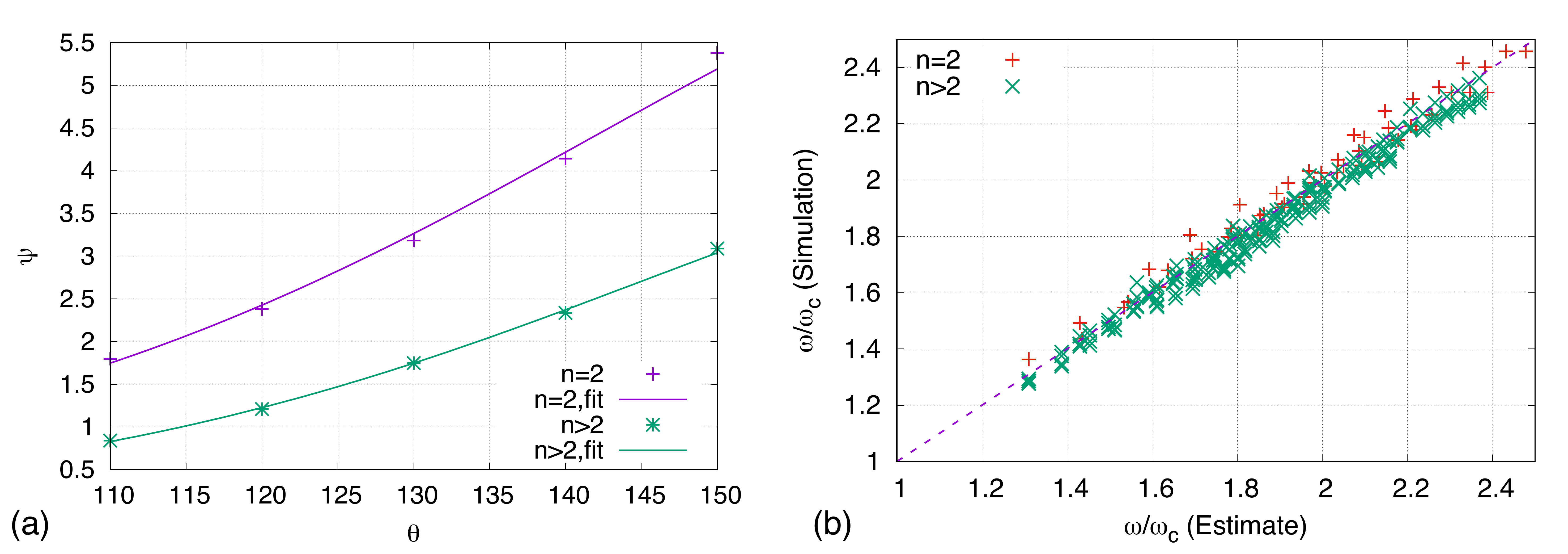}}
 \caption{(a) The fitted values for the parameter $\psi$ as a function of $\theta$ (degree) for $\ang{110}<\theta< \ang{150}$. (b) Comparison of frequencies predicted by the model (Eqs.\ \eqr{omegan}--\eqr{psi_fit}) and the simulation results.  } 
 \label{fig:freq_mode2-6_scaling}
\end{figure}

Nevertheless, for small $Bo$ and large $\theta$, \eg, small drops on hydrophobic surfaces, $\omega_n/\omega_{n,Ra}$ for $n\ge 3$ is almost independent of $n$ and the scaling relation can be used to estimate the frequencies for high-order oscillation modes, 
\begin{equation}
	\omega_n = \omega_{n,Ra} \beta(\theta, Bo) = \omega_c \sqrt{(n-1)n(n+2)} \beta(\theta, Bo)
	\label{eq:omegan}
\end{equation}
where $\beta = \omega_n/\omega_{n,Ra}$  is a function of $\theta$ and $Bo$ and can be further estimated as 
\begin{equation}
	\beta(\theta, Bo) =  \beta_0(\theta)\big(1+\psi (\theta) Bo\big)^{1/4}
	\label{eq:beta}
\end{equation}
where $\beta_0=\beta(\theta, Bo=0)$ depends on $\theta$ only, the simulation results of which are shown in Fig.\ \ref{fig:freq_mode2-6_theta}(b). The effect of $Bo$ is incorporated by the correction function, the form of which is taken to be similar to that for the first mode (Eq.\ \eqr{correction_finiteBo}). Tests have been made for all $\theta$, indicating that Eq.\ \eqr{beta} is a good approximation for $\ang{110} \le \theta \le \ang{150}$. The coefficient $\psi$ as a function of $\theta$ for $n=2$ and other $n>2$ modes are obtained by fitting the simulation results, see Fig.\ \ref{fig:freq_mode2-6_Bo}(c). The fitting errors are all smaller than about 4\%. The values of $\psi$ for $n=2$ and $n>2$ modes are given in Fig.\ \ref{fig:freq_mode2-6_scaling}(a). 

Finally, for convenience of using the model Eqs.\ \eqr{omegan}--\eqr{beta}, fitting expressions for $\beta_0(\theta)$ and $\psi(\theta)$ are provided. Expressions similar to Eq.\ \eqr{eta_fit} are employed, with the exponential correction term dropped, since only the high $\theta$ cases are considered in this model,   
\begin{align}
	\log(\beta_0(\theta)) & = j_0+j_1 (1+\cos\theta) +\left[\exp\left(\frac{(1+\cos\theta) ^{j_2} }{j_3} \right)-1\right]\,,
	\label{eq:beta_0_fit}\\
	\log(\psi(\theta)) & = k_0+k_1 (1+\cos\theta)\,.
	\label{eq:psi_fit}
\end{align}
{The fitting constants for $\beta_0$ are $[j_0,j_1,j_2,j_3]=[0.153,0.877, 6.509,94.6]$}, and those for $\psi$ are $[k_0,k_1]=[1.93,-2.08]$ for $n=2$ and $[1.44,-2.48]$ for $n>2$. The fitted functions, Eqs.\ \eqr{beta_0_fit} and \eqr{psi_fit} are plotted in Fig.\ \ref{fig:freq_mode2-6_theta}(b). The estimated frequencies for the high-order modes based on the model (Eqs.\ \eqr{omegan} and \eqr{beta}) with fitting correlations (Eqs.\ \eqr{beta_0_fit} and \eqr{psi_fit}) are compared with the simulation results in Fig.\ \ref{fig:freq_mode2-6_scaling}(b) and a good agreement is observed. 

\section{Conclusions}
\label{sec:conclusions}
The axisymmetric modes for the natural oscillation of sessile drops on flat surfaces with free contact lines (FCL) have been investigated in the present study through a combined numerical-theoretical approach. The FCL condition represents a limiting case for the contact line mobility, for which the contact line can move freely with the equilibrium contact angle. The simulation approach is validated by the oscillations of a free drop and a sessile drop with the contact angle $\theta=\ang{90}$. In total, 121 cases have been studied to cover a wide range of contact angles ($\ang{50}\le \theta \le \ang{150}$) and Bond numbers ($0\le Bo \le 0.88$). Particular attention is paid to the first mode, since it is usually the dominant mode. An inviscid theoretical model has been developed for the first mode based on conservation of the total energy (the sum of surface, potential, and kinetic energy). Based on the assumption that the shapes of the drop during oscillation are similar to the equilibrium shapes under different body forces, the surface energy can be expressed as function of the centroid location. Eventually the model yields an explicit expression of the first-mode frequency as a function of $\theta$ and $Bo$, with all model parameters fully determined by the equilibrium drop theory and the simulation results. The model predictions of first-mode frequencies agree well with the simulation results for all $\theta$ and $Bo$. In particular, the predicted frequencies match almost perfectly with simulation results for $Bo=0$. To investigate the high-order modes ($n>1$), a different initial condition was utilized to initiate drop oscillations, with initial perturbation of surface energy biased toward the high-order modes. The simulation results show that the variations of frequencies over $\theta$ and $Bo$ for the first and high-order modes are similar: the frequencies normalized by capillary frequency $\omega/\omega_c$ decrease with $\theta$ and increase with $Bo$. The rate of increase of $\omega/\omega_c$ over $Bo$ is reduced as $\theta$ decreases. For the high-order modes, the frequencies $\omega_n$ scale with the corresponding Rayleigh frequencies $\omega_{n,Ra}$. This scaling relation performs better for large $\theta$ and small $Bo$ since the equilibrium shape is closer to a sphere. A simple model is proposed to predict the frequencies of high-order modes for $\ang{110} \le \theta \le \ang{150} $. A good agreement between the model predictions and the simulation results is observed. 

\section*{Acknowledgement}
This work was supported by the startup fund at Baylor University and the grant (\#1853193) from the National Science Foundation. The Baylor High Performance and Research Computing Services (HPRCS) have provided the computational resources that have contributed to the research results reported in this paper. The authors also acknowledge Dr.~St\'ephane Popinet for the contribution to the development of the  \emph{Basilisk} code. We also thank Dr.~Jorge Alvarado and Dr.~Joshua Bostwick for helpful discussions. 

\appendix
\section{Equilibrium sessile drop on flat surfaces}
\label{sec:equilibrium}
\begin{figure}
 \centering{\includegraphics[width=0.35\textwidth]{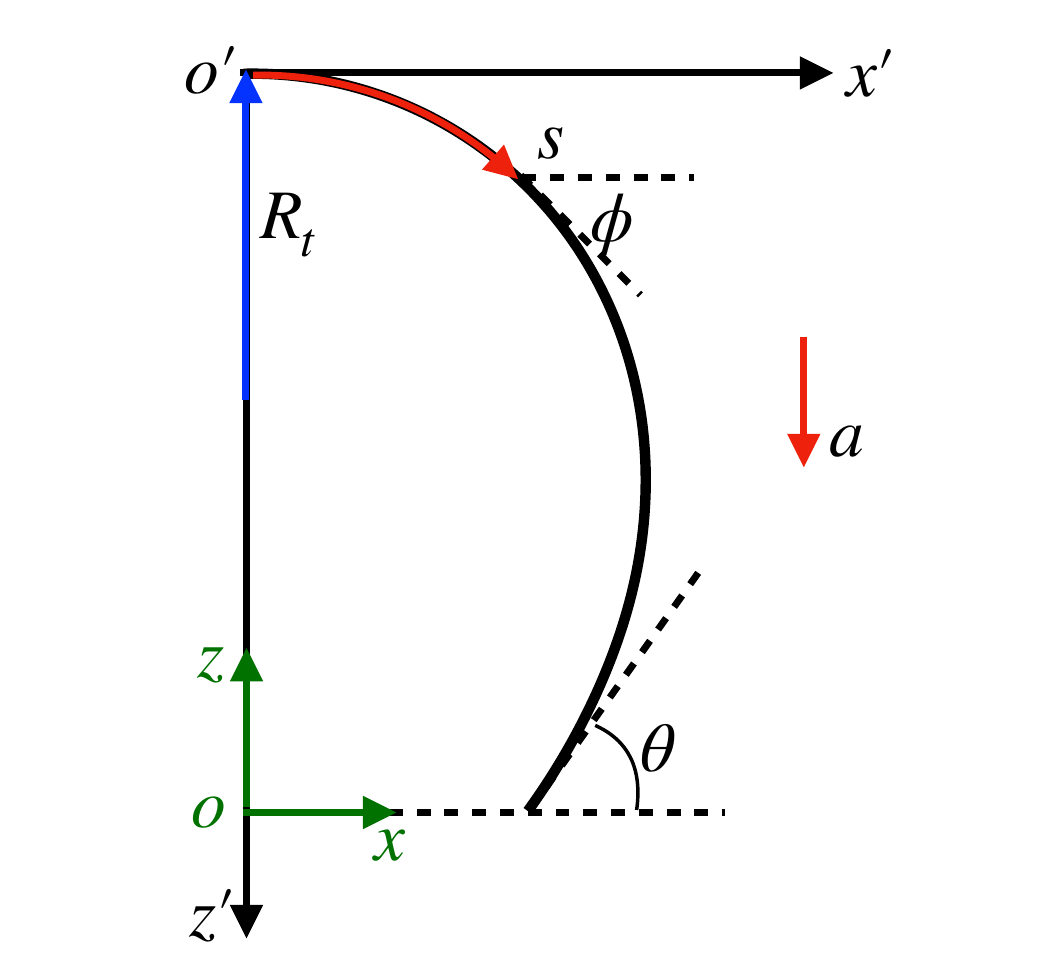}}
 \caption{Sketch of the axisymmetric quasi-static sessile drop profile.}
 \label{fig:SessileDropLayout}
\end{figure}

When a sessile drop is at its equilibrium state, the Laplace pressure is in balance with the hydrostatic pressure, namely
\begin{equation}
  \sigma (\frac{1}{{R}_1} + \frac{1}{{R}_2}) = \frac{2 \sigma}{R_t} + (\rho_l - \rho_g) g z',
  \label{eq:sessile_equilibrium}
\end{equation}
 where $\rho_l$ and $\rho_g$ are the water and air densities, respectively. For convenience, we transform the coordinate system to $(x',z')$ as shown in Fig. \ref{fig:SessileDropLayout}. The radius of curvature at the top of the sessile drop is denoted by $R_t$. The two principal radii of curvature, ${R}_1$ and ${R}_2$, can be calculated as
\begin{equation}
  \frac{1}{{R}_1} = \frac{\delta \phi}{\delta s}, \frac{1}{{R}_2} = \frac{\sin \phi}{x'},
  \label{eq:radii_of_curvature}
\end{equation}
 where $s$ is the curvilinear coordinate along the drop surface starting from the top of the drop, see Fig. \ref{fig:SessileDropLayout}. Combining Eqs.\ \eqr{sessile_equilibrium} and \eqr{radii_of_curvature}, it yields
\begin{equation}
  \frac{\delta \phi}{\delta s} = 2 \kappa_t - \frac{(\rho_l - \rho_g) g z'}{\sigma} - \frac{\sin \phi}{x'},
  \label{eq:sessile_equilibrium_2}
\end{equation}

Furthermore, it can also be shown from geometry that
\begin{equation}
  \frac{\delta x'}{\delta s} = \cos \phi\,,\quad  \frac{\delta z'}{\delta s} = \sin \phi. 
  \label{eq:dzds}
\end{equation}
For a given $R_t$, Eqs.\ \eqr{sessile_equilibrium_2} and \eqr{dzds} can be integrated numerically from $\phi=0$ to $\theta$ to obtain the shape of the equilibrium sessile drop. Since $R_t$ is not known a priori, iteration is required to identify the value of $R_t$ that will yield a drop volume matching $V_d$.

\section{Surface area over centroid location for equilibrium sessile drop}
\label{sec:equilibrium_varyg}
\begin{figure}[tbp]
 \centering
 \includegraphics[width=0.5\textwidth]{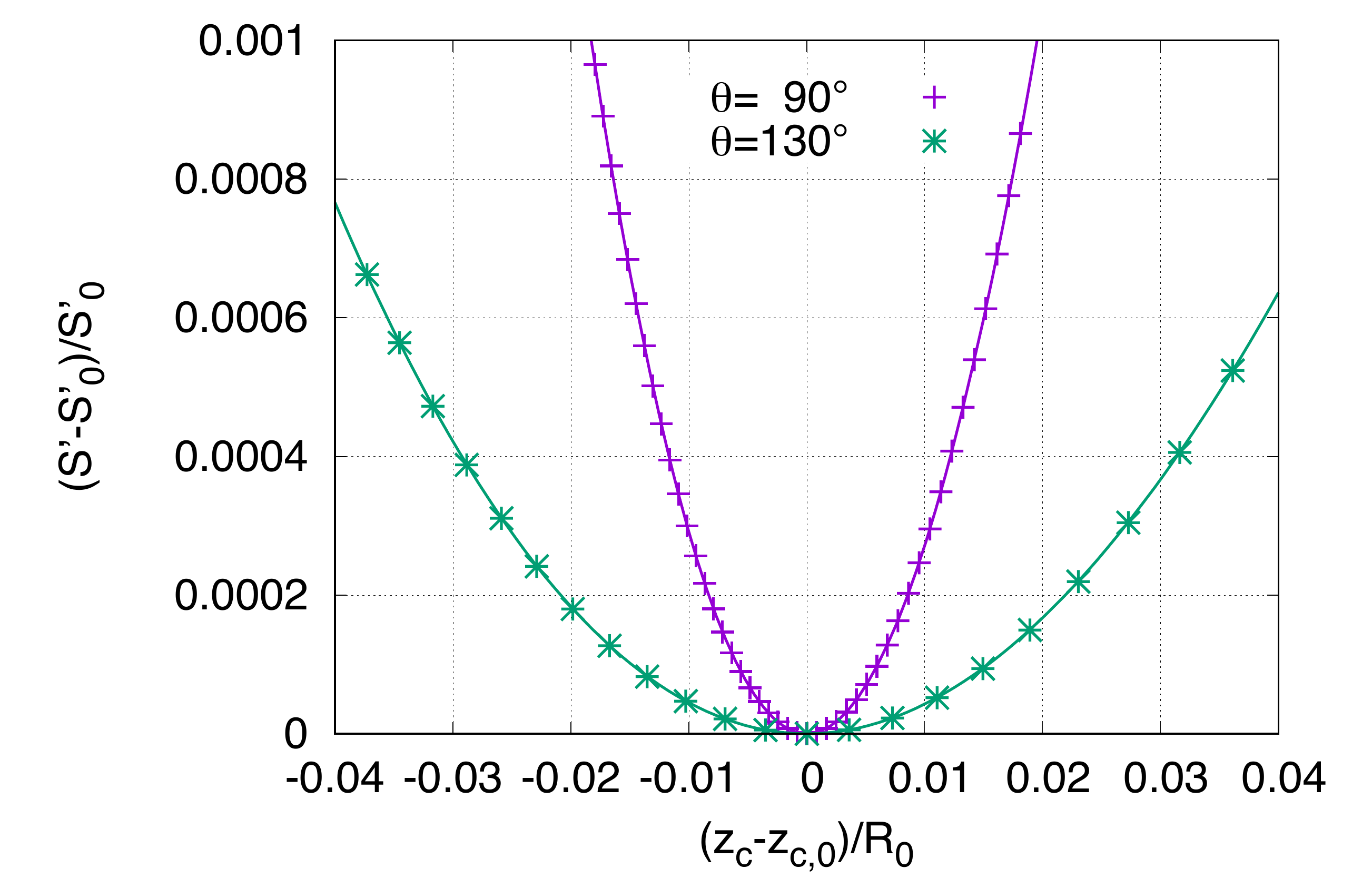}
 \caption{Variation of the modified surface area $dS'/S_0'$ for the equilibrium shape of the drop under different gravities with respect to the centroid location $dz_c/R_0$ for $\theta =$ (a) $50$\textdegree\ and (b) $150$\textdegree. }
 \label{fig:equil_shape_g}
\end{figure}

When a body force acceleration $a$ is applied to a sessile drop along the $z'$ (see Fig. \ref{fig:SessileDropLayout}), the equilibrium shape of the drop will vary. When $a>0$, the equilibrium centroid position will decrease, i.e.  $z_c<z_{c,0}$, or vice versa. Regardless of whether $a$ is positive or negative, the total surface energy characterized by the modified drop surface area $S'$ (Eq.\ \eqr{modified_drop_surf_area}) will increase. The deviation of the modified surface area from the value for $a=0$, $S'-S'_0$, is a function of the deviation of $z_c$ from $z_{c,0}$. The variation of $S'-S'_0$ over $z_c-z_{c,0}$ for $\theta=\ang{90}$ and $\ang{130}$ are shown in Fig.\ \ref{fig:equil_shape_g}. A series expansion near the origin with terms higher than third order truncated yields Eq.\ \eqr{FindingHTheta}. The parameters $\eta$ and $\xi$ for a given $\theta$ can be obtained by fitting the results of the equilibrium sessile drop theory for the range $-0.088 < \rho_l a R_d^2 / \sigma <0.088$, see the solid lines in Fig.\ \ref{fig:equil_shape_g}. The fitting errors for all cases are less than 0.5\%, indicating the truncation error in Eq.\ \eqr{FindingHTheta} is small. Furthermore, the values of $\eta$ and $\xi$ are not sensitive to the fitting range of $a$, as along as $|\rho_l a R_d^2 / \sigma|<0.1$. 
 
\section{Effect of Volume of the Drop}
\label{sec:AppendixVarVol}
Sessile drops of different volumes were simulated to confirm that the effect of drop volume on the normalized frequency $\omega/\omega_c$ is small in the present problem. Four different volume-based radii $R_d=1$ 2.5, 5, and 10 mm are considered, and the corresponding $Oh$ for these drops are 0.01429, 0.00571, 0.00286, and 0.00143, respectively. The contact angle is $\ang{90}$ and $Bo=0$. It is shown that for the range of $Oh$ tested, the frequencies for different modes are independent of $Oh$. As expected, the first-mode frequencies for all $V_d$ or $Oh$ match the second-mode Rayleigh frequencies for the corresponding free drops. 

\begin{figure}[tbp]
 \centering
 \includegraphics[width=0.5\textwidth]{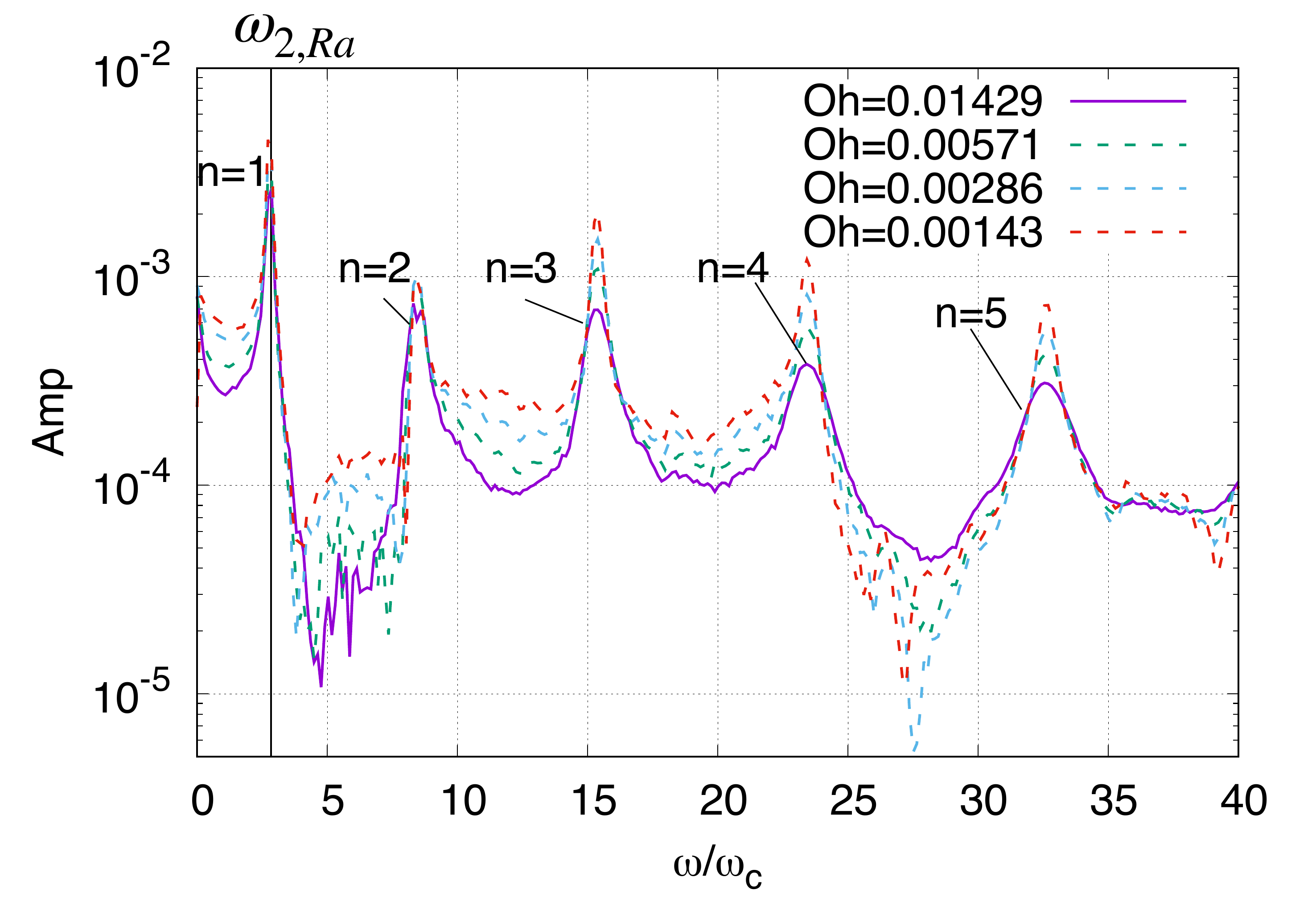}
 \caption{Frequency spectrum for the drop height for $Oh$ = 0.01429, 0.00571, 0.00286, and 0.00143. The IC2 approach is used to initiate the drop shape oscillation, which activates the first and also high-order modes. }
 \label{fig:modeIC2_VarVol}
\end{figure}

\section{Velocity boundary condition for free contact line on flat surfaces}
\label{sec:BC_CL}
The contact-line velocity boundary conditions (BC) for mobile contact lines were derived by Bostwick and Steen \cite{Bostwick_2014a}. The linearized Young-Dupr\'e equation for flat solid surfaces can be written as 
\begin{equation}
	\pd{u_n}{s} + \left(\kappa_c \cot \theta \right) u_n = 0\, . 
	\label{eq:BC_FCL_flat_vel}
\end{equation}
where $u_n$ is the contact-line velocity component normal to the drop surface and $\kappa_c$ is the radius of curvature of the drop surface at the contact line location. For FCL (referred to natural contact line in their paper), Bostwick and Steen further assumed that $\kappa_c$ is fixed at its equilibrium value, \ie, 
\begin{equation}
	\kappa_c=\sin \theta\,,
	\label{eq:kappa}
\end{equation}
with the radius of the contact line at the equilibrium state, $x_{cl}$, as the reference length scale, then Eq.\ \eqr{BC_FCL_flat_vel} reduces to 
\begin{equation}
	\pd{u_n}{s} + \cos \theta u_n = 0\, , 
	\label{eq:BC}
\end{equation}
which is the BC used in their model (Eq.~(3.8) in the paper \cite{Bostwick_2014a}). However, when the sessile drop with FCL oscillates, the contact line is perturbed from its equilibrium position and $\kappa_c$ also changes accordingly. Therefore, Eq.\ \eqr{kappa} is not strictly valid  in general, and Eqs.\ \eqr{kappa} and \eqr{BC} will introduce a small error in the contact-line velocity BC. The only exception is that, when $\theta=\ang{90}$, both of Eqs.\ \eqr{BC_FCL_flat_vel} and \eqr{BC}  reduce to $\pd{u_n}{s} =0$, then the two BC's are equivalent. This explains why the oscillation frequencies predicted by Bostwick and Steen's model agree with the present results very well for $\theta=\ang{90}$ and some deviations have been observed for large $|\theta-\ang{90}|$, see Figs.\ \ref{fig:freq_mode1_Bo0} and \ref{fig:freq_mode2-6_theta}.

%

\end{document}